# Accretion and differentiation of the terrestrial planets with implications for the compositions of early-formed Solar System bodies and accretion of water

D.C. Rubie[1*], S.A. Jacobson[1,2], A. Morbidelli[2], D.P. O'Brien[3], E.D. Young[4], J. de Vries[1], F. Nimmo[5], H. Palme[6], D.J. Frost[1]

[1]Bayerisches Geoinstitut, University of Bayreuth, D-95490 Bayreuth, Germany (dave.rubie@uni-bayreuth.de)

[2]Observatoire de la Cote d'Azur, Nice, France

[3]Planetary Science Institute, Tucson, Arizona, USA

[4]Dept. of Earth and Space Sciences, UCLA, Los Angeles, USA

[5] Dept. of Earth & Planetary Sciences, UC Santa Cruz, USA

[6] Forschungsinstitut und Naturmuseum Senckenberg, Frankfurt am Main, Germany

* Corresponding author

**Abstract.**

In order to test accretion simulations as well as planetary differentiation scenarios, we have integrated a multistage core-mantle differentiation model with N-body accretion simulations. Impacts between embryos and planetesimals are considered to result in magma ocean formation and episodes of core formation. The core formation model combines rigorous chemical mass balance with metal-silicate element partitioning data and requires that the bulk compositions of all starting embryos and planetesimals are defined as a function of their heliocentric distances of origin. To do this, we assume that non-volatile elements are present in Solar System (CI) relative abundances in all bodies and that oxygen and $H_2O$ contents are the main compositional variables. The primary constraint on the combined model is the composition of the Earth's primitive mantle. In addition, we aim to reproduce the composition of the Martian mantle and the mass fractions of the metallic cores of Earth and Mars. The model is refined by least squares minimization with up to five fitting parameters that consist of the metal-silicate equilibrium pressure and 1-4 parameters that define the starting compositions of primitive bodies. This integrated model has been applied to six Grand Tack N-body accretion simulations. Investigations of a broad parameter space indicate that: (1) accretion of Earth was heterogeneous, (2) metal-silicate equilibration pressures increase as accretion progresses and are, on average, 60-70% of core-mantle boundary pressures at the time of each impact, and (3) a large fraction (70-100%) of the metal of impactor cores equilibrates with a small fraction of the silicate mantles of proto-planets during each core formation event. Results are highly sensitive to the compositional model for the primitive starting bodies and several accretion/core-formation models can thus be excluded. Acceptable fits to the Earth's mantle composition are obtained only when bodies that originated close to the Sun, at <0.9-1.2 AU, are highly reduced and those from beyond this distance are increasingly oxidized. Reasonable concentrations of $H_2O$ in Earth's mantle are obtained when bodies originating from beyond 6-7 AU contain 20 wt% water ice (icy bodies that originated between the snow line and this distance did not contribute to Earth's accretion because they were swept up by Jupiter and Saturn). In the six models examined, water is added to the Earth mainly after 60-80% of its final mass has accreted. The compositional evolution of the mantles of Venus and Mars are also constrained by the model. The FeO content of the Martian mantle depends critically on the heliocentric distance at which the Mars-forming embryo originated. Finally, the Earth's core is predicted to contain 8-9 wt% silicon, 2-4 wt% oxygen and 10-60 ppm hydrogen, whereas the Martian core is predicted to contain low concentrations (<1 wt%) of Si and O.

# 1. Introduction

It is widely recognized that the four terrestrial planets of our Solar System, Mercury, Venus, Earth and Mars, formed over a time period on the order of up to 100 million years (e.g. Jacobson et al., 2014). According to current astrophysical theories, these planets formed in three stages: (1) Planetesimal formation. During this stage, dust condensed from the solar nebula, a disk of gas and dust rotating around the growing Sun, settled to the midplane of the disk and coagulated to form km- to multi-km-size solid planetesimals. (2) Planetary embryo formation. In a disk of planetesimals, the largest bodies grow fastest due to gravitationally enhanced mutual collisions creating a population of embryos (Greenberg et al., 1978). By the end of this stage, a population of oligarchic planetary embryos, lunar to Mars-mass objects, has emerged amongst the swarm of remnant planetesimals (Kokubo & Ida, 1998). (3) Planet formation. The final stage of accretion was dominated by the mutual gravitational interactions of the embryos and was characterized by large, violent collisions (e.g. Benz et al., 1989). In recent years, there has been significant progress in modelling this giant impact stage. For example, the numerical *N-body accretion* simulations of O'Brien et al. (2006), starting with embryos and planetesimals initially located in the region between 0.3 and 4 astronomical units (AU), resulted often in the formation of 3-4 planets that provide a reasonable match to many characteristics of the terrestrial planets in the Solar System. The resulting planets are generally distributed between 0.5 and 2 AU and have masses ranging from 0.3-1.6 $M_e$ (where $M_e$ = mass of the Earth).

An outstanding problem in modelling the formation of the terrestrial planets has been to reproduce the small mass of Mars (~0.1$M_e$). Classical simulations, in which the giant planets form on or near their current orbits (i.e. no giant planet migration), have generally produced a Mars-like planet that is too massive and located too far from the Sun. An important insight into this problem has been made based on the idea of Hansen (2009). A small Mars can form if the planetesimal disk becomes truncated at 1-1.5 AU. The Grand Tack model of Walsh et al. (2011) achieves this truncation by invoking the large-scale radial gas-driven migration of Jupiter and Saturn. First, they migrate inwards towards the Sun during the early stages of Saturn's growth (such radial migration is common for giant planets discovered around other stars). Once Saturn achieved a critical mass and was in resonance with Jupiter, these planets then migrated back outwards. This inward-then-outward migration (the "Grand Tack") truncated the disk of planetesimals and planetary embryos at around 1 AU and subsequent accretion in this truncated disk then results in a system of planets matching the terrestrial planets of the Solar System including, in particular, a Mars-like planet with a correct mass.

At present, the main constraints on the validity of accretion models are the masses and orbital characteristics of the final planets in comparison with the actual terrestrial planets of the Solar System. On the other hand, considerable information is available concerning the chemistry of the silicate mantles of the Earth and, to a lesser extent, Mars, Venus and Mercury (e.g. McDonough, 2003; Palme and O'Neill, 2013; Dreibus and Wänke, 1985; McSween, 2003; Taylor, 2013; Righter and Chabot, 2011; Treiman 2009; Robinson and Taylor, 2001; Taylor and Scott, 2003). Combining core-mantle differentiation with accretion modelling should provide important new constraints on both processes.

There have been no attempts to comprehensively integrate geochemical evolution models with the N-body accretion simulations. Although several studies have considered the accretion of water-bearing material onto the terrestrial planets (e.g. Morbidelli et al., 2000; O'Brien et al., 2014), only two studies have attempted to address the bulk composition of accreted material and its implications for the final compositions of the planets (Bond et al., 2010; Elser et al., 2012). In these studies, predictions of equilibrium solar-nebula condensation models were combined with the results of N-body simulations in order to estimate the bulk chemistry of planets as a function of their final locations (heliocentric distances) in the Solar System. There are a number of problems with this approach. In particular, no account is taken of core-mantle differentiation even though this has a large influence on the chemistry of planetary mantles - on which the observational constraints are based.

In this study we combine N-body accretion and core-mantle differentiation models following the approach of Rubie et al. (2011). We thus model differentiation and geochemical evolution of all the terrestrial planets simultaneously. Such an approach provides an important additional test of the viability of N-body simulations. In addition, models of planetary core formation can be significantly improved and refined. Note that the aim is not to define the exact conditions and processes of accretion and differentiation – which is clearly impossible for such stochastic and complex events. Instead we aim to provide robust indications of the likely ranges of conditions and processes.

Here we describe the methodological approach in detail and discuss results from six Grand Tack accretion simulations. In a subsequent paper (Jacobson et al., in preparation) the method described here will be applied to a large number of accretion simulations in order to examine the results statistically.

## 2. Methodology

### 2.1. N-body simulations

We derived the accretion history of each planet from six Grand Tack N-body simulations. Each simulation run is integrated with Symba (Duncan et al., 1998) tracking both embryos, which interact with all particles in the simulation, and planetesimals, which interact only with the Sun, embryos and the giant planets (for a review of similar simulations see Morbidelli et al., 2012). These simulations assume perfect accretion and are run for 150 to 200 Myr with the effects of gas included for the first 0.6 Myr. Since these are Grand Tack simulations, the giant planets do not begin near their current orbits and they migrate during the gas phase of the disk. Here we follow the prescription set out in the supplementary material of Walsh et al. (2011) in the section "Saturn's core growing in the 2:3 resonance with Jupiter". In brief, Jupiter and Saturn begin at 3.5 and 4.5 AU on circular orbits, respectively, then over 0.1 Myr migrate inward to 1.5 and 2 AU. Meanwhile, Saturn's mass grows linearly from 10 $M_e$ to its current mass, and when it reaches a mass close to its current mass the migration physics in the resonance changes and the giant planets then migrate outwards (Masset & Snellgrove, 2001; Morbidelli & Crida, 2007; Pierens & Nelson, 2008). They continue to migrate outwards as the gas in the disk dissipates exponentially over the next 0.5 Myr. Once the gas is gone, they are then stranded on orbits with semi-major axes of 5.25 and 7 AU, respectively. These locations are appropriate for a late giant planet instability at ~500 Myr (Nice model; Morbidelli et al., 2007) that will place them and the outer ice giants (Neptune and Uranus) on their current orbits (Levison et al., 2011). The treatment of the nebula gas and the giant planets are identical in each of the six simulations.

Only the initial locations and masses of the embryos and planetesimals distinguish the six simulations. All simulations begin with an inner disk of embryos and planetesimals and an outer disk of planetesimals, which is disturbed inward by the outward migration of the giant planets. Table 1 lists the parameters of these disks for each simulation (the parameters for the inner planetesimal disk are listed above those for the outer planetesimal disk). The naming convention of the simulations is designed to convey the following information: (a) an "i" indicates that the mass of the embryos increases as a function of initial semi-major axis across the range given in column 4 of Table 1, (b) the ratio of the total mass of the embryos to planetesimals (4:1 or 8:1), (c) the initial mass of the largest embryo (0.025, 0.05 or 0.08 $M_e$), and (d) the run number in the simulation suite with those disk parameters.

The six simulations were chosen from a larger suite of simulations (Jacobson and Morbidelli, 2014) on the basis of the geochemical approach taken here. For the latter, it is important to only use simulations that produce planets at heliocentric distances and with masses that are both similar to those of the Solar System planets. Of course, in practice, it is impossible to reproduce all the characteristics of the Solar System simultaneously. The results of core-mantle differentiation depend on the pressures of metal-silicate equilibration and therefore on the mass of accreting bodies. Because we want to compare these simulations to one another, all starting embryo and planetesimals masses are scaled by a factor such that each final Earth analogue has a mass of exactly 1 $M_e$. This mass adjustment factor ranges from 0.91 to 1.06 (Table 1). Each simulation produced an Earth and a Venus analogue, as shown in Fig. 1. In most of the chosen simulations, the Earth-like planet experiences a late (>50 Myr) giant impact (Fig. 2; Jacobson et al., 2014). Four of the simulations have only one Mars analogue and two have multiple Mars analogues. Simulation i-4:1-0.8-6 also results in a stranded embryo at ~2.5 AU. This body is unlikely to be stable because of its proximity to Jupiter and is therefore not considered here. We do not discuss Mercury in this paper because none of the selected simulations produced Mercury analogues.

*2.2. Core-mantle differentiation*

Embryos grow through collisions with other embyros and with planetesimals. Because of the high energies involved, impacts inevitably result in extensive melting and magma ocean formation, especially in the case of giant impacts (Tonks and Melosh, 1993; Rubie et al., 2007; de Vries et al., 2014). Deep magma ocean formation enables liquid Fe-rich metal to chemically equilibrate with silicate liquid at high pressure before segregating to the core. Unless impacting bodies are fully oxidized or the metal becomes oxidized in the magma ocean (see below), each impact results in an episode of core formation in the target body and core formation is therefore multistage. By modeling the compositions of equilibrated metal and silicate for every impact event, we track and store the evolving mantle and core compositions of all embryos and proto-planets in the N-body simulation as accretion proceeds (Rubie et al. 2011). Note that in addition to the giant impacts that affect the Earth-like planets, the embryos that impact the proto-Earth have themselves also experienced giant impacts. All such events are modeled here.

In most studies of core formation in the Earth, assumptions are made regarding oxygen fugacity and how it evolves because this is a critical parameter that controls how siderophile (metal-loving) elements partition between metal and silicate (e.g. Gessmann et al., 1999; Wade and Wood, 2005; Siebert et al., 2013). Instead of making such assumptions, we follow the alternative approach of Rubie et al. (2011) and assign bulk compositions to the original embryos and planetesimals. We consider here only the non-volatile elements Al, Mg, Ca, Fe, Si, Ni, Co, Nb, Ta, V and Cr and assume that these elements have Solar System bulk composition ratios, (i.e. CI chondritic) but with concentrations that are enhanced relative to CI by 22% for the refractory elements Ca, Al, Nb, Ta, etc. and by 11% for V (Table S1, Supplementary Data). Enhanced concentrations of refractory elements are based mainly on the Earth's mantle having an Al/Mg ratio that is significantly higher than the CI Al/Mg ratio (see Rubie et al., 2011, for a detailed discussion). The enhancement of refractory element compositions is important when modelling the bulk composition of the Earth's mantle correctly. Of the above elements, Al, Mg and Ca are lithophile and do not partition into metal whereas Fe, Si, Ni, Co, Nb, Ta, V and Cr are weakly to moderately siderophile and partition to varying degrees into metal (e.g. Kegler et al., 2008; Mann et al., 2009; Siebert et al., 2012, 2013). W and Mo are not included here because the partitioning of these elements depends on the sulphur content of the metal (Wade et al., 2012), and sulphur, as a volatile element, is not included in the current model. The oxygen content of the bulk composition is a critical parameter which is constrained by least squares refinements as described in detail below. The oxygen content is fixed by specifying the fraction of total Fe that is present as metal (as opposed to that present as FeO in silicate); this fraction can be varied from 0.999 to 0 (giving compositions that range from highly reduced to fully oxidized). The oxygen content in highly-reduced compositions can be decreased further by specifying that a fraction of the total Si is also dissolved initially in metal, for which there is evidence from enstatite chondrites (Keil, 1968), although not from iron meteorites (which may therefore have originated from relatively oxidized bodies).

### 2.2.1. Element partitioning

The extents to which siderophile elements partition between metal and silicate determine the core and mantle compositions of planetary bodies during and after core formation and are dependent on the pressure ($P$), temperature ($T$) and oxygen fugacity ($f_{O2}$) conditions of metal-silicate equilibration. Partitioning of a given element i is described by the partition coefficient

$$D_i^{met-sil} = \frac{C_i^{met}}{C_i^{sil}} \qquad (1)$$

where $C_i^{met}$ and $C_i^{sil}$ are the wt% or molar concentrations of i in metal and silicate respectively. The partition coefficient $D_i^{met-sil}$ is dependent on oxygen fugacity, in addition to pressure and temperature (e.g. Mann et al., 2009). Thus the distribution coefficient $K_D$, which is the partition coefficient normalized to the partition coefficient of Fe (raised to the power $n/2$), thus making it independent of $f_{O2}$, is used here:

$$K_D = \frac{D_i^{met-sil}}{(D_{Fe}^{met-sil})^{n/2}} = \frac{X_i^{metal} \left[ X_{FeO}^{silicate} \right]^{n/2}}{X_{iO_{n/2}}^{silicate} \left[ X_{Fe}^{metal} \right]^{n/2}}. \qquad (2)$$

Here $X$ represents the mole fractions of components in metal or silicate liquids and $n$ is the valence of element i. The dependence of $K_D$ on $P$ and $T$ is described by:

$$\log_{10} K_D = a + b/T + c\, P/T. \qquad (3)$$

The parameters $a$, $b$ and $c$ are constants and the values used here, their uncertainties and sources are listed in Table S2 of the Supplementary Data. Some effects of using alternative $K_D$ models based on other studies are also discussed in the Supplementary Data.

In the case of Si partitioning, we use the relationship:

$$\log K_D(\text{Si}) = a + b/T + (c_1\, P + c_2\, P^2 + c_3\, P^3)/T \qquad (4)$$

according to which the pressure effect is weak and becomes even weaker as pressure increases (Mann et al., 2009; Rubie et al., 2011); the parameter values used are listed in Table S2 of the Supplementary Data.

$K_D$ can also be affected by the compositions of the silicate liquid, especially in the case of high-valence elements such as W and Mo (Wade et al., 2012), and the composition of the liquid metal (Tuff et al., 2011). However, for the elements considered here, such compositional dependencies are small and are not taken into account in this study.

Oxygen partitioning, which is critical for solving the mass balance described below, is described by a thermodynamic model that is consistent with experimental data obtained up to 70 GPa and 3500 K (Frost et al., 2010; Rubie et al., 2011). As $K_D$ for oxygen depends on the compositions of the phases involved, its value is determined from the thermodynamic model by an iterative procedure.

### 2.2.2. Mass balance

The major-element molar compositions of coexisting chemically-equilibrated silicate and metal liquids depend on how elements partition between the two phases and are therefore a function of pressure and temperature. The compositions are expressed as:

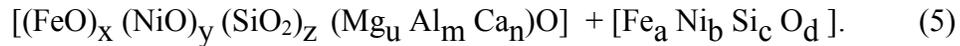

$$[(\text{FeO})_X\ (\text{NiO})_y\ (\text{SiO}_2)_Z\ (\text{Mg}_u\ \text{Al}_m\ \text{Ca}_n)\text{O}]\ + [\text{Fe}_a\ \text{Ni}_b\ \text{Si}_c\ \text{O}_d\,]. \qquad (5)$$

<div style="text-align:center"><em>silicate liquid</em>         <em>metal liquid</em></div>

The indices u, m and n are determined solely from the bulk composition because Mg, Al and Ca are lithophile elements that do not partition into the metal. This leaves seven unknown indices (x, y, z, a, b, c and d) for the elements that do partition into metal: these seven indices are determined by mass balance combined with element partitioning data. Four mass balance equations (for Fe, Si, Ni and O) are used together with two expressions for the metal-silicate partitioning of Si and Ni and the thermodynamic model for oxygen partitioning at high pressure (Frost et al., 2010). The seven indices are determined by an iterative procedure as described in detail by Rubie et al. (2011, Supplementary Data). Oxygen fugacity ($f_{O2}$), which is a critical parameter that controls partitioning, is not specified but can be determined from the resulting Fe content of the metal and the FeO content of the silicate (Rubie et al., 2011, Fig. 2). The trace elements (Co, Nb, Ta, V and Cr) do not influence the mass balance significantly because of their low concentrations and their respective concentrations in metal and silicate are determined by element partitioning data alone.

When the mass balance/partitioning approach is applied to the equilibration of impactors that contain only a small fraction of metal (due to a highly oxidized bulk composition), the "metal" liquid is sometimes predicted to have such a high oxygen content that it has to be regarded as an oxide liquid. In this case, we add the oxidised material to the silicate liquid and there is no core formation event.

An important process that controls the mass balance involves the dissolution of Si into metal by the reaction:

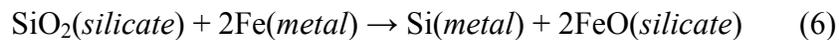

$$SiO_2(silicate) + 2Fe(metal) \rightarrow Si(metal) + 2FeO(silicate) \qquad (6)$$

(Javoy, 1995; Rubie et al., 2011). This reaction shows that every mole of Si that dissolves into metal causes two moles of FeO to be transferred to the silicate. Reaction 6 is therefore a very effective mechanism for increasing the oxygen fugacity and FeO content of the mantle during core formation (Rubie et al., 2011). This reaction, combined with the accretion of relatively oxidized material (Rubie et al., 2011), provides a good explanation for the empirical oxygen fugacity increases in the continuous core formation model of Wade and Wood (2005). In contrast, the "self-oxidation" or "oxygen pump" mechanism proposed by these authors cannot produce the necessary increase in mantle FeO content during core formation (see section 154.3.2.2.3 in Rubie et al., 2014b).

*2.3. Compositions of early-formed Solar System bodies*

In order to apply the above mass balance/element partitioning approach to modelling planetary differentiation, it is necessary to assign a bulk composition (i.e. with a specific oxygen content) to every initial embryo and planetesimal in the N-body simulations over heliocentric distance intervals as great as 0.5 to 10 AU. To do this, we make the basic assumption that oxygen content varies systematically with heliocentric distance, rather than randomly, due, for example, to the systematic temperature structure of the solar nebula (e.g. Wood, 2000). We then consider three composition-distance models in order to explore the broadest possible parameter space. (1) The compositions (i.e. oxygen contents) of all starting bodies were identical from 0.5 to 10 AU across the early Solar System. (2) Bodies that formed relatively close to the Sun were oxidized and those originating at greater heliocentric distances were reduced. (3) Bodies that formed close to the Sun were reduced and those from greater heliocentric distances were oxidized. As shown below, we can categorically exclude the first two of these three models.

*2.4. P-T conditions of metal-silicate equilibration*

As shown by numerous studies of core formation in the Earth (e.g. Table 3 of Rubie et al., 2008, 2014b), the pressure-temperature conditions of metal-silicate equilibration are critical for the resulting mantle and core compositions because of the dependences of $K_D$ on $P$ and $T$ for the various siderophile elements. Here we follow the simple approach used by Rubie et al. (2011) for modelling multistage core formation. For each impact-induced core formation event we assume that the metal-silicate equilibration pressure $P_e$ is a constant fraction of the target's evolving core-mantle boundary pressure:

$$P_e = f_P \times P_{CMB} \qquad (7)$$

where $f_P$ is a constant proportionality factor. $P_{CMB}$, the core-mantle boundary (CMB) pressure at the time of impact, is calculated using Equation 2.73 in Turcotte and Schubert (2002) using the approximations that (a) core density = 2.5 × mantle density and (b) average mantle density scales linearly with planetary mass, as defined by the masses and densities of Earth and Mars. It is assumed, for simplicity, that $f_P$ is constant throughout accretion and is independent of the mass and velocity of colliding bodies. The parameter $f_P$ is thus considered to represent the average of a range of values that, in reality, depend on the mass of impactors, impact velocities and impact angles (de Vries et al., 2014). The validity of assuming a constant value of $f_P$ is discussed below in section 7.1.

Similar to the approach of previous studies (e.g. Wade and Wood, 2005; Rubie et al., 2011) we assume that the metal-silicate equilibration temperature, $T_e$, lies between the peridotite liquidus and solidus at the equilibration pressure $P_e$. We use a combination of liquidus/solidus temperatures of Herzberg and Zhang (1996), Trønnes and Frost (2002) and the "FeO corrected" liquidus and eutectic temperatures of Liebske and Frost (2012, Fig. 7) at higher pressures (>24 GPa). The melting temperatures of Liebske and Frost (2012) are based on a thermodynamic model; their mantle liquidus temperatures are close to those determined experimentally by Andrault et al. (2010) but lie at temperatures that are several 100 K below the liquidus temperatures of Fiquet et al. (2010). The model used here for the metal-silicate equilibration temperature is

$$P < 24.0: \qquad T_e = 1874 + 55.43 P - 1.74 P^2 + 0.0193 P^3 \qquad (8a)$$

$$P \geq 24 \text{ GPa}: \qquad T_e = 1249 + 58.28 P - 0.395 P^2 + 0.0011 P^3 \qquad (8b)$$

(see Fig. S1, Supplementary Data). There are uncertainties of at least ±100 K on peridotite melting temperatures, even at moderate pressures, as shown by inconsistencies between the studies cited above (see also Nomura et al., 2014). We therefore consider below the effects of adjusting the equilibration temperatures given by Eq. 8 upwards or downwards by adding or subtracting a temperature adjustment $\Delta T$.

## 2.5. Core-mantle differentiation of primitive bodies

Primitive metal-bearing bodies are assumed to differentiate early, before they are involved in their first impact event, due to the heat produced by the decay of $^{26}$Al. To model this, we use the mass balance/partitioning approach described above and assume complete core-mantle equilibration at a pressure given by Eq. 7. In order to apply Eq. 7, we calculate a fictive core-mantle boundary pressure based on the mass of the body, as described above. The results of such initial differentiation events affects the final results of the accretion/core formation model only when the metal of the impactor's core only equilibrates partially in the magma ocean (see section 3.1).

## 3. Extent of metal-silicate equilibration

In most models of core formation in the Earth, it has generally been assumed implicitly that complete chemical equilibration occurs between all Fe-rich metal and the entire silicate magma ocean (e.g. Wade and Wood, 2005). There are actually two separate issues to consider here (Morishima et al., 2013): (a) the fraction of an impactor's core that equilibrates in the magma ocean and (b) the proportion of the target's magma ocean/mantle that equilibrates with metal.

### 3.1. Extent of metal equilibration

When a differentiated body impacts a proto-planet, the metallic core of the impactor is expected to plunge through the magma ocean to eventually merge with the core of the target body. The extent to which the plunging core equilibrates chemically depends on the extent to which it breaks up and emulsifies into stable droplets with a radius on the order of a centimetre (Rubie et al., 2003). Although the cores of small impactors may emulsify and equilibrate completely, the behaviour of the cores of large "giant" impactors remains uncertain (Dahl and Stevenson, 2010; Rubie et al., 2011; Samuel, 2012; Deguen et al., 2014).

The degree to which metal equilibrates chemically in a magma ocean affects siderophile element concentrations (Rubie et al., 2011) and is of major importance for interpreting W isotopic anomalies in terms of the timing of planetary core formation (e.g. Nimmo and Agnor, 2006; Nimmo et al., 2010; Rudge et al., 2010). Below we start with the assumption that all metal equilibrates and then examine later the consequences of this not being the case (section 7.2).

### 3.2. Extent of mantle/magma ocean equilibration

An accretional impact on an embryo results in the creation of a semi-spherical melt pool that may extend to the embryo's core-mantle boundary in the case of a Moon-forming giant impact (Tonks and Melosh, 1993, Fig. 13). This immediately excludes the possibility that the impactor's core could equilibrate with the entire mantle of the embryo (see also Sasaki and Abe, 2007; Ćuk and Stewart, 2012). Furthermore, Deguen et al. (2011) argued that an impactor's core will sink through the melt pool together with turbulently-entrained silicate liquid in a descending plume (Fig. 3). The plume expands with increasing depth as more silicate liquid becomes entrained; thus the volume fraction ($\phi_{met}$) of metal in the metal-silicate plume decreases with depth and is given by:

$$\phi_{met} = \left(\frac{r_0}{r}\right) = \left(1 + \frac{\alpha z}{r_0}\right)^{-3}. \tag{9}$$

Here $r_0$ is the initial radius of the impactor's core, $r$ is the radius of the plume, $z$ is depth in the magma ocean and $\alpha$ is a constant with a value ~0.25 (Fig. 3; Deguen et al., 2011). We use this model to estimate the mass fraction of the embryo's mantle that is entrained, as silicate liquid, when the descending metal-silicate plume reaches the bottom of the magma ocean as defined by $f_P$ (see Eq. 7). This fraction of the silicate mantle then defines the molar amounts of silicate liquid components that equilibrate chemically with the metal of the impactor's core, as formulated by (5) above. Following chemical equilibration, the resulting metal is added to the proto-core and the equilibrated silicate is mixed with the fraction of the mantle that did not equilibrate to produce a compositionally homogeneous mantle. Such mixing is the consequence of vigorous convection in the early Earth.

## 4. Model constraints and fitting parameters

Our primary aim is to determine whether combined accretion/core formation models can produce an Earth-like planet with a mantle composition that is close to that of the Earth's primitive mantle. In order to match the Earth's mantle composition, we perform least squares regressions with nine constraints as provided by the primitive mantle concentrations of the elements Fe, Si, Ni, Co, Nb, V, Ta and Cr (Palme and O'Neill, 2013) and the tightly-constrained Nb/Ta ratio of 14.0 (±0.3) (Münker et al., 2003). An important assumption is that the primitive mantle concentrations listed by Palme and O'Neill (2013) are representative of the entire mantle. Arguments in support of this assumption (on which all core-formation models are based) can be found, for example, in Drake and Righter (2002), Rubie et al. (2011) and Palme and O'Neill (2013).

A secondary aim is to determine if other terrestrial planets can be produced in the same simulations with realistic mantle compositions. In detail, the latter are poorly known compared with the composition of the Earth's mantle. The most reliable data are for the FeO content of the Martian mantle which has been estimated to be ~18 wt% (Dreibus and Wänke, 1985, 1987; McSween, 2003). Righter and Chabot (2011) and Taylor (2013) have also estimated the concentrations of a number of siderophile elements in the Martian mantle based on the SNC meteorites. The internal composition of Venus is particularly poorly-known but based on similarities between basalts on the two planets it may be similar to that of the Earth's interior (Treiman, 2009). Because of the uncertainties involved and for additional reasons discussed below in the case of Mars, these compositions are not used as constraints when performing least squares regressions. An additional parameter to consider is the mass fraction of each planet's core, which is calculated using the mass balance approach adopted here. The actual values are 0.32 for the Earth and an estimate of 0.15-0.23 for Mars (Sohl and Spohn, 1997; McSween, 2003).

We initially use 2-5 fitting parameters to obtain fits to the Earth's mantle composition. One parameter that is of crucial importance is the proportionality factor $f_P$ that defines the metal-silicate equilibration pressure as a fraction of the core-mantle boundary pressure at the time of impact (Eq. 7). The other 1-4 parameters specify the oxygen contents of the bulk compositions of the starting embryos and planetesimals and the initial distribution of these compositions in the early Solar System, as described in the next section. There are several additional parameters, including the metal-silicate equilibrium temperature $T_e$ (Eq. 8) and the fraction of impacter cores that equilibrate, which are fixed in the initial refinements: the effects of varying these are discussed below.

In all cases we minimize the sum of the squares of the weighted residuals

$$\chi^2 = \Sigma \left( \left( C_C^i - C_M^i \right) \middle/ \sigma_M^i \right)^2 \qquad (10)$$

where $C_C^i$ is the calculated Earth mantle wt% concentration of element $i$, $C_M^i$ is the actual wt% concentration and $\sigma_M^i$ is the uncertainty on $C_M^i$. We report reduced chi squared values

$$\chi^2_{red} = \frac{\chi^2}{\upsilon} \qquad\qquad (11)$$

where $\upsilon$ is the number of degrees of freedom, given by $\upsilon = N\text{-}n\text{-}1$, with $N$ being the number of observations and $n$ the number of fit parameters.

Least squares regressions are performed with the simplex method (Nelder and Mead, 1965) using the "amoeba" code from Press et al. (2007). This involves making initial guesses in the form of $n+1$ sets of fit parameters, where $n$ is the total number of fit parameters. This enables an extremely broad parameter space to be explored rapidly and efficiently. The regressions are re-run several times, each time using revised guesses, in order to avoid local minima.

## 5. Results

We now explore the validity of three different models for the initial bulk compositions of primitive embryos and planetesimals as a function of their heliocentric distances of origin.

### 5.1. Constant composition-distance model (homogeneous accretion)

First we test the possibility that all initial embryos and planetesimals had the same composition, which means that accretion was homogeneous for all planets. Assuming, as described above, that non-volatile siderophile elements were present in Solar-System (CI) relative proportions, the only compositional variable is the oxygen content. This is fixed by defining the proportion of Fe initially present as metal (with the remainder being present as FeO in silicate). This proportion, $X^{met}_{Fe}$, is one fitting parameter (Fig. 4a). The other fitting parameter is $f_p$ (Eq. 7) which defines metal-silicate equilibration pressures during accretion. The result of a least squares regression of these two parameters is shown as the "best-fit homogeneous accretion" model in Figs. 4a and 5 and Table 2. The parameter space investigated in this regression covers a broad range of values ($X^{met}_{Fe}$: 0.20-0.99 and $f_p$ : 0.2 to 1.0) that converge to $X^{met}_{Fe} = 0.84\text{-}0.85$ and $f_p = 0.72\text{-}0.80$ with $\chi^2_{red} = 67\text{-}74$ (Table 2). The results for all six simulations show that, although resulting

Earth mantle concentrations for FeO, Ni and Co are almost perfect, $SiO_2$, V and Cr concentrations and the Nb/Ta ratio are much too high (Table 2, Fig. 5). These results emphasize the importance of considering as many elements as possible when constraining core formation models. Rubie et al. (2011) obtained a similar result and concluded that it is not possible to simultaneously match the mantle concentrations of $SiO_2$ and FeO when accretion is homogeneous; instead accreting material needs to include both reduced and oxidized compositions. Additional problems are that (1) the calculated FeO concentrations of the mantles of all Venus and Mars analogues are virtually identical (7.87-8.39 wt%) and are indistinguishable from that of the Earth's mantle and (2) the calculated mass fraction of the Martian core is 0.27 in all simulations. These results are totally inconsistent with current estimates (see above) and this model can therefore be excluded.

It has been proposed that core formation occurred under oxidising conditions with the silicate magma ocean initially having a high FeO content of ~20 wt% (Rubie et al., 2004; Siebert et al., 2013). According to this hypothesis, oxygen partitions (as FeO) into the metallic core under oxidizing conditions, which is supposed to result in a final Earth mantle FeO concentration of ~8 wt%, with several wt% oxygen in the core. We have attempted, unsuccessfully, to reproduce this result by fixing $X_{Fe}^{met} = 0.55$ (Fig. 4a), which gives an initial silicate FeO concentration of ~20 wt%, and refining only $f_p$ over the range 0.2 to 1.0. This results in an extremely poor fit for almost every element with $f_p = 0.64$-1.0 and $\chi_{red}^2 = 4937$-5801 (model "core formation under oxidizing conditions" in Fig. 5 and Table 2). The fundamental problem with this hypothesis is that it fails to take account of the effects of the partitioning of Si into the core which becomes significant at high temperatures even under oxidizing conditions. By mass balance, this increases the mantle FeO concentration (Eq. 6) and almost completely counteracts the effect of oxygen partitioning. Thus the FeO content of Earth's mantle is only reduced from 20 wt% to 17-18 wt% (rather than to 8 wt%) when Si partitioning is included (see also Frost et al., 2008) and is the main cause of the extremely high value of $\chi_{red}^2$. As with the "best fit homogeneous accretion" model, all resulting planets have very similar mantle FeO contents (in this case 17-20 wt%) and, in addition, the predicted mass fraction of the Earth's core of 0.21-0.23 is much too low.

The compositional evolution of the Earth's mantle during accretion, based on these two homogeneous accretion models, is shown in Fig. 5 for simulation 4:1-0.5-8. All six simulations show very similar trends although the exact paths for Ni and Co show some variation, depending on the individual impact histories.

When comparing the results of the two homogeneous accretion models (Fig. 5), it is noteworthy that under the more oxidizing conditions of the second model discussed above, the concentrations of the siderophile elements Fe, Ni and Co in the mantle are relatively high whereas the mantle concentrations of the more lithophile elements Si, V, Nb and Ta are relatively low. This difference is caused by the mass fraction of the mantle of the "core formation under oxidizing conditions" model being larger (Table 1), due to oxidation of Fe, which causes the lithophile elements to be more diluted.

*5.2. Variable composition-distance models (heterogeneous accretion)*

Because the homogeneous accretion models described in the previous section give very poor results, we now consider heterogeneous accretion. Following Rubie et al. (2011), the simplest approach is to consider two compositions, with different oxygen contents that are described by the parameters $X_{Fe}^{met}(1)$ and $X_{Fe}^{met}(2)$, where $X_{Fe}^{met}$ is the fraction of total Fe present as metal (as above). The parameter $X_{Fe}^{met}(1)$ describes the oxygen content of bulk compositions in the innermost Solar System whereas $X_{Fe}^{met}(2)$ gives the oxygen content in the outermost Solar System. An additional parameter, $\delta(1)$, is the heliocentric distance separating the respective regions in which the two compositions formed (Fig. 4b-c).

In order to fully explore the broadest possible parameter space, we first consider a model in which the innermost compositions are relatively oxidised and the outmost compositions are reduced (Fig. 4b). Although it is difficult to find a plausible justification for such a model, it can be considered as a modification of the model of core formation under oxidising conditions. In this case, material that accretes early would be oxidized and material that accretes later would be reduced and would thus help to achieve a reduction in the Earth's mantle FeO content (which is not possible when accretion is homogeneous, as shown above). There are thus four fitting parameters: $f_p$, $X_{Fe}^{met}(1)$, $X_{Fe}^{met}(2)$ and $\delta(1)$. The ranges of these parameters explored in the fit are $f_p$: 0.2 to 1.0; $X_{Fe}^{met}(1)$: 0.15 to 0.6; $X_{Fe}^{met}(2)$: 0.62 to 1.0; $\delta(1)$: 0.5 to 4.0 AU. We have obtained two types of results from the least squares fits. In some simulations, the refined value of $\delta(1)$ becomes less than the heliocentric distance (0.5 or 0.7 AU) of the innermost primitive bodies and the model then becomes indistinguishable from the "best fit homogeneous accretion" model discussed above. When the refined value of $\delta(1)$ remains greater than the minimum heliocentric distance of primitive bodies, the fit is comparable to, but slightly worse than, those of the "best fit homogeneous accretion" models. This model can also be excluded.

The best results of Rubie et al. (2011) for the Earth's mantle were obtained when early-accreted material was highly reduced and later-accreted material was more oxidized (see also Wade and Wood, 2005; Wood et al., 2006). This is also consistent with early accreted material being volatile-depleted and later accreted material being volatile-enriched (Schönbächler et al., 2010). We therefore fit a step model in which innermost compositions are highly reduced and compositions from further out are partially oxidized. We initially used four fitting parameters as above and explored the following parameter space $f_p$: 0.2 to 1.0; $X_{Fe}^{met}(1)$: 0.62 to 1.0; $X_{Fe}^{met}(2)$: 0.15 to 0.6; $\delta(1)$: 0.5 to 4.0 AU. The resulting fits give $\chi_{red}^2 \approx 70$. However, the fits were improved enormously by fixing $X_{Fe}^{met}(1) = 0.999$ and allowing a fraction of the available Si (refined to $X_{Si}^{met} = 0.11 - 0.14$) to be dissolved initially in the metal of the reduced composition, resulting in $\chi_{red}^2$ =0.82-3.0 for the six simulations, with the other refined parameter values as listed in Table 3. For the Earth's mantle, the fits are excellent for all elements, especially when the propagated uncertainties on the final calculated compositions are considered (Table 3). The results for Mars are highly variable and show a distinct dichotomy, with the final mantle FeO content being either very low (0.2-6 wt%) or very high (18-29 wt%) (Table 3). These values depend on whether the Mars-forming embryo forms in the region of highly-reduced compositions or in the region of oxidized compositions. In addition, when Mars has an unrealistically long accretion time (e.g. simulation 4:1-0.5-8) the composition of an originally highly-reduced embryo can be strongly modified through a collision with an oxidised embryo.

The "step" composition-distance model of Fig. 4c is in no way unique and comparable (or even better) results are obtained, for example, by defining a gradient in the oxygen content of early-formed bodies (Rubie et al., 2014a). As discussed below, a "gradient" model may be more realistic than the "step" model and we therefore further develop the former.

## 6. Oxidation in the inner Solar System and accretion of Earth's water

Based on the results present above, the bulk compositions of embryos and planetesimals forming at heliocentric distances less than 1.1 to 1.7 AU were initially highly reduced - with essentially all Fe being present as metal and with 6-7 wt% Si dissolved in early-formed metallic cores. According to the results of the successful step model presented above (Fig. 4c), initial core formation in such bodies occurred at oxygen fugacities ~4 log units below the iron-wüstite buffer (IW-4). Such a low oxygen fugacity is far below the intrinsic oxygen fugacities recorded by both the terrestrial planets at the end of core formation (Frost et al., 2008) and most meteorite parent bodies. Bodies that formed beyond 1.1 to 1.7 AU were, in contrast, considerably more oxidized (Fig. 4c).

Oxygen fugacities of a solar gas are ~4 to 5 orders of magnitude more reducing than the intrinsic oxygen fugacities at which the terrestrial planets and most meteorite parent bodies formed ($\sim IW-2$) but are consistent with the region of highly-reduced compositions at <1.1 to 1.7 AU postulated here. Thus extensive oxidation must have occurred in the early solar system with the main consequence being the oxidation of Fe to form a FeO component in silicates. The most likely oxidant was water. A possible explanation for the trends in oxidation state shown in Fig. 4c is as follows: Due to the net inflow of material in the solar nebula, ice-covered dust moves inwards from beyond the snow line. Inside the snow line, water ice sublimes, thus adding $H_2O$ to the vapor phase (Cuzzi and Zahnle, 2004). As temperatures continue to rise and material continues to move inwards, $H_2O$-rich vapor reacts with Fe-bearing dust which results in oxidation. Inward still, vapor is $H_2O$-poor because the products of sublimed water ice have not mixed all the way to the innermost solar system (Cuzzi and Zahnle, 2004). Here Fe remains free of oxidation and highly-reduced compositions are preserved. Spitzer and Herschel data for the TW Hydra protoplanetary disk suggest the presence of a ring of water vapour at about 4 AU (Zhang et al. 2013), suggesting in turn, that water is concentrated near the snow line and becomes less abundant inwards towards the star.

However, there is a problem with the above explanation and the precise mechanism for elevating oxygen fugacity from solar to values recorded in planetary materials by water remains unclear. Based on the equilibrium (Krot et al., 2000)

$$H_2 + 1/2\,O_2 \Leftrightarrow H_2O \qquad\qquad (12)$$

$$\log(f_{O_2}) \approx 2\log\left(\frac{H_2O}{H_2}\right) + 5.67 - \frac{256664}{T(K)} \qquad (13)$$

an increase in oxygen fugacity by 5 log units requires $H_2O/H_2$ to increase by a factor of 375. Simple geometric arguments for a protoplanetary disk with $\sigma \propto R^{-1}$, where $\sigma$ is surface density of gas and $R$ is radial distance from the Sun, limits the enhancement of water to a factor of 25 if all of the solid ice from 5 to 50 AU were concentrated in an annulus of 2 AU inside of a snowline at 5 AU. Concentration from a maximum disk radius of at least $\sim$ 300 AU, rather than 50 AU, is required for the necessary enrichment of water. Grossman et al. (2012) emphasized the difficulty of stabilizing the FeO component in silicates by nebular processes and proposed two solutions to the problem of achieving sufficient oxidation. Either oxidation occurred primarily by aqueous processing within planetesimals or, alternatively, as the result of kinetic suppression of metallic nickel-iron nucleation due to surface tension.

Carbonaceous chondrite meteorites (CI, CM) may serve as models for the oxidized, $H_2O$-bearing bodies that are incorporated into the model below. These rocks exhibit tell-tale signs of extensive water-rock reactions in the form of secondary phyllosilicates, carbonates and oxides (e.g., Kerridge and Bunch, 1979; Krot et al., 1998; McSween et al. 2003). Models for the hydrology of the parent bodies of these rocks indicate that the parent bodies had excesses of water relative to that required to form hydrated minerals and that large reservoirs of water were retained within them (Young et al. 1999; Young 2001; Palguta et al. 2010). There is growing evidence that many asteroids may still possess significant fractions of water to this day (Hsieh and Jewitt, 2006; Jewitt et al., 2009; Kuppers et al., 2014). The density of Ceres, for example, is consistent with a water content of up to 50 wt%.

We now develop a composition-distance model that may be more realistic than that shown in Fig. 3c. We thus propose a compositional gradient model to include fully-oxidized $H_2O$-bearing compositions, as represented by C-type asteroids and CI chondrites (Fig. 6a). We also include the possibility of a second outer oxidation gradient (Fig. 6a) but emphasize that this geometry is not unique. However, the broad variation in oxygen contents of Fig. 6 is consistent with the range of oxidation states recorded by meteorites (e.g. from highly-reduced enstatite chondrites to fully-oxidized CI chondrites), which is not the case with the step model of Fig. 4c. The introduction of an oxidation gradient also overcomes the dichotomy in calculated Martian mantle FeO concentrations obtained in the step model, as described above. First we discuss how we treat the incorporation of water into the planets during accretion and core formation.

*6.1. Incorporation of water into the terrestrial planets*

A composition-distance model for primitive embryos and planetesimals that includes $H_2O$-bearing bodies and is consistent with the oxidation mechanism described above is shown in Fig. 6a. To be consistent with CI chondrite compositions, the water-bearing bodies are fully oxidized and contain 20 wt% $H_2O$. Because such bodies contain no metal, there is no core formation event when they collide with and accrete to a proto-planet. The fully-oxidized material is therefore mixed homogeneously into the mantle of the proto-planet, which means that a planetary mantle becomes $H_2O$-bearing as soon as the first C-type asteroid accretes. When a subsequent core-formation event occurs, due to the accretion of a metal-bearing body, the $H_2O$ in the fraction of the mantle that equilibrates with the metal of the impactor is considered to dissociate. The hydrogen content of the equilibrating silicate is assumed to partition into the molten metal (Okuchi, 1997) and/or is lost to space by hydrodynamic escape. The oxygen released by the dissociation of $H_2O$ then contributes to oxidation of the silicate mantle (e.g. Ringwood, 1984; Wade and Wood, 2005; Sharp et al., 2013).

The distance $\delta(3)$, beyond which planetesimals are $H_2O$-bearing in the composition-distance model of Fig. 6, is not treated as a fitting parameter in least squares regressions but is simply adjusted to obtain approximately 1000 ppm $H_2O$ in the Earth's primitive mantle at the end of accretion (O'Neill and Palme, 1998; Palme and O'Neill, 2013), although this concentration might be a slight overestimate (Hirschmann, 2006, pers. comm.). In addition, we assume here that all dissociated hydrogen partitions into the liquid metal that merges with the core; this assumption then enables an upper limit to be placed on the hydrogen content of the cores of the terrestrial planets. Note that the values of $\delta(3)$ obtained below (6.0-6.9 AU) are consistent with the origin of the asteroid belt according to the Grand Tack hypothesis (Walsh et al., 2011; O'Brien et al., 2014) but, as discussed below, should not be confused with the location of the original snow line.

### 6.2. Evolution of planetary chemistry during accretion

The results of best-fit models for the six Grand Tack simulations, based on the composition-distance gradient model of Fig. 6a, are detailed in Tables 4-7 and Fig. 6b-d. The chemical evolutions of the mantles of the resulting Earth-like planets are shown in Fig. 7. As an example, the complete set of chemical evolution results for the mantles and cores of all three planets in simulation i-4:1-0.8-6 are presented in Table S3 (Supplementary Data). The evolution of oxygen fugacity in these models is also described in the Supplementary Data and is documented in Table S3. In the composition-distance models (Fig. 6), primitive bodies originating at <0.92-1.16 AU have highly reduced compositions and compositions beyond this distance become increasingly oxidised and are almost completely oxidised at and beyond 1.8-2.8 AU ($X_{Fe}^{met}(2)$ refines to values ranging from $3 \times 10^5$ to 0.2 but not to zero). There appears to be no tendency for $\delta(1)$ and $\delta(2)$ to converge to the same value – which suggests that the gradient model provides better fits than the step model of Fig. 4c. Metal-silicate equilibration pressures lie in the range $0.58$-$0.72 \times P_{CMB}$ at the time of each impact.

Water-bearing bodies originate beyond 6.0-6.9 AU, which results in 810-1070 ppm $H_2O$ in the Earth's mantle. Most water is added to the Earth after 60-80% of its total mass has accreted. However in one simulation (4:1-0.5-8) there is a large increase in water content after 33% of its mass has accreted due to the addition of a series of $H_2O$-bearing oxidized planetesimals (Fig. 7a).

Simple model temperatures and pressures at the midplane of the solar accretion disk predict that the snow line migrated inward from about 5 AU to 2 AU as the rate of accretion of the protoplanetary disk onto the central star decreased from $10^{-7}$ $M_e$ yr$^{-1}$ to $10^{-8}$ $M_e$ yr$^{-1}$ (Armitage, 2011). Observations of other protoplanetary disks support the expectation that the position of the water snow line was approximately at these distances from the Sun (e.g., Zhang et al., 2013). The snow line marks an increase in solid surface density so its initial position may have facilitated the formation of Jupiter near 5 to 6 AU. The region between the giant planets from ~3 to 6 AU was sparsely populated by icy bodies due to the accretion of the giant planets, but the Grand Tack scenario naturally explains the addition of distantly (>6 AU) accreted icy planetesimals during the outward migration phase. These distant icy planetesimals were hypothesized to be the C-complex asteroids in the Main Belt (Walsh et al., 2011), which is consistent with our interpretation of them here.

The similarity of refined parameters in the six Grand Tack simulations and the low values of $\chi^2_{red}$ (Table 4) are a strong indication of the robustness of the model. The variations between parameters values are generally small and are indicative of the uncertainties involved in the models. The evolutions of element concentrations in the Earth's mantle are mostly similar for all six simulations although in detail there are differences caused by different impact histories that result from the stochastic nature of accretion (Fig. 7). In particular, simulation 8:1-0.8-8 is different because the original embryo that grows to form the Earth has a relatively oxidized composition ((Fig. 7a).

Based on the results of the six simulations, the Earth's core is predicted to contain 2-4 wt% oxygen, 8-9 wt% silicon and 9-58 ppm hydrogen (the likely presence of 2 wt% sulphur in the core (McDonough, 2003) is not considered here). Our estimated concentrations of O and Si in the Earth's core are consistent with estimates based on geochemical arguments (Allègre et al. 1995), partitioning experiments (Ricolleau et al., 2011; Tsuno et al., 2013) and equation of state determinations (Fischer et al., 2011, 2014a). The low oxygen content, relative to silicon, is also consistent with the results of shockwave data on the Fe-S-O system (Huang et al., 2012). However, other sound velocity studies (Badro et al., 2007) and ab initio simulations (Alfè at al., 2002, 2007) have predicted that the outer core contains a significantly higher concentration of O and a lower concentration of Si than the values estimated here. The calculated mass fraction of the core is only slightly lower than the actual value, which suggests that loss of silicate material by collisional erosion has not been significant (c.f. O'Neill and Palme, 2008).

Mantle and core compositions for the Venus-like planets in the six simulations are listed in Table 6. In general, the mantle compositions are similar to that of the Earth's mantle, which is consistent with the conclusions of Treiman (2009) based on similarities between basalts on the two planets. In particular, the mantle FeO concentrations lie mostly in the range 7-8 wt%, although two values are significantly lower.

The six Grand Tack simulations result in nine Mars-like planets with masses that range from 0.032 to 0.28 $M_e$ (Figs. 1 and 2). The calculated FeO contents of the mantles of these planets are highly variable and range from 4.86 to 23.8 wt%. Four of these bodies have an extended accretion history lasting for >100 Myr that involves impacts with other embryos (simulations 4:1-0.5-8, 4:1-0.25-7, 8:1-0.25-2 and i-4:1-0.8-4 – see Figs. 2 and 6). Such results are inconsistent with Mars being a stranded embryo with a short accretion time of <10 Myr (Dauphas and Pourmand, 2011) and are not further considered here.

The calculated mantle and core compositions of the five Mars-like bodies that accrete on short (<10 Myr) timescales are listed in Table 7 together with the Martian mantle composition determined by Taylor (2013) and Righter and Chabot (2011). The initial locations and oxidation states of the embryos are shown, as solid symbols, in Fig. 6. The final mantle FeO concentrations of four of these bodies are significantly less than the actual value of 18 wt% and the fifth value of 20 wt% (simulation i-4:1-0.8-6) is close but is perhaps slightly too high. The Martian oxidation state (mantle FeO, Ni and Co contents) is strongly dependent on the heliocentric distance at which the embryo forms with respect to the proposed oxidation gradient in the early solar system, as shown in Fig. 6. As for FeO, the mantle Ni and Co concentrations are too high in simulation i-4:1-0.8-6 but are too low in the other four Mars-like planets. The differences in oxidation state also control the mass fraction of the Martian core which is predicted correctly only by simulation i-4:1-0.8-6 (Table 7). The calculated $H_2O$ contents of Martian mantles vary from 0 to 4770 ppm (Table 7) and depend on the individual accretion histories.

*6.3 Effect of accreted water on evolving mantle chemistry*

The final water content of the Earth's mantle can be changed simply by varying the value of parameter $\delta(3)$ in Fig. 6. For example, if the value is changed from 6.9 to 6.0 AU in simulation 4:1-0.25-7, the final mantle $H_2O$ content changes from 1000 ppm to 3270 ppm and the H content of the core increases from 28 ppm to 96 ppm. It is important to emphasize, however, that the exact value of $\delta(3)$ is dependent on the initial distribution of planetesimals in the N-body simulations and how far they extend from the Sun. For example, if planetesimals initially extend out to 13 AU (compared with 9.5 AU in the current simulations), $\delta(3) \approx 8.7$ AU is required to produce ~1000 ppm $H_2O$ in the Earth's mantle.

It has been proposed that water that is accreted to the Earth is the main cause of mantle oxidation through the degassing of $H_2$ (Sharp et al., 2013). Because the accretion of water and loss of $H_2$ are modelled here, we can test the magnitude of this proposed oxidation effect by comparing the results of the 4:1-0.25-7 accretion/core formation model with (a) zero $H_2O$ in primitive bodies, (b) 20 wt% $H_2O$ in bodies located initially at >6.9 AU and (c) 20 wt% $H_2O$ in bodies located initially at >6.0 AU. These three starting conditions result in final mantle $H_2O$ concentrations of 0 ppm, 1000 ppm and 3270 ppm, respectively. Least squares regressions, however, produce almost identical results, with fitted parameter values and mantle concentrations varying by <1%, irrespective of whether or not accreting bodies contain $H_2O$. For example, mantle Ni concentrations of 1808, 1821 and 1813 ppm are obtained for the above three cases, respectively. We conclude therefore that evolving oxidation state during planetary accretion is controlled by the oxygen content of accreting material (Fig. 6) and the partitioning of Si into the core (Eq. 6) and that the effects of accreted water are negligible. This is because (1) the amounts of $H_2O$-bearing accreted material are small (Fig. 1, Table 5) and (2) the proportion of the target's mantle that is involved in metal-silicate equilibration events is also small (Fig. 3). Of course, as discussed above, the oxidation state of primitive bodies is likely to be caused by the effects of $H_2O$: however, this results from oxidation that occurred prior to the main stage of planetary accretion.

# 7. Effects of varying other parameters

We now consider the effects of varying several parameters that were held constant when calculating the results presented above.

## 7.1 Metal-silicate equilibration pressures for cores of embryos and planetesimals

The depth of melting during planetary accretion depends strongly on the mass of impacting bodies (e.g. Rubie et al., 2008; de Vries et al., 2014). Whereas embryos in the N-body simulations considered here have masses in the range 0.02-0.08 $M_e$, the planetesimals masses are orders of magnitude smaller (Table 1). Furthermore, each "planetesimal" in the N-body simulations is actually a tracer that consists of a swarm of much smaller bodies, e.g. ~100 km in diameter (O'Brien et al., 2006). Thus, if the lifetime of magma oceans was short, the metallic cores of impacting planetesimals may have equilibrated at much lower pressures than those of embryos, in contradiction to the assumption of section 2.4 that $f_P$ is independent of the mass of impacting bodies.

In order to test this possibility, we have refined the core formation model in simulation 4:1-0.25-7 by fixing the metal-silicate equilibration pressures for embryo impacts with $f_P$ values in the range 0.7-0.95 (Eq. 7) and have determined that best fit values of $f_P$ for planetesimals impacts then lie in the range 0.45-0.5 (Fig. 8). Significantly lower values (e.g. 0.2) result in high values of $\chi^2_{red}$ (e.g. >10). Good fits are obtained when $f_P$ for embryos lies in the range 0.7-0.75. However, $\chi^2_{red}$ is almost indistinguishable from the case where $f_P$ values are assumed to be the identical for all impacting bodies (red symbol in Fig. 8), which therefore justifies the simplified approach taken in this paper.

Based on these preliminary results, average high pressures of metal-silicate equilibration for small (e.g. 100 km diameter) impacting bodies are consistent with deep global magma oceans having a long life time due to the presence of a dense insulating atmosphere (e.g. Abe, 1993)

*7.2 Extent of metal equilibration*

The results presented in previous sections are based on the assumption that the metal of impactor cores equilibrates completely with silicate liquid (e.g. due to complete emulsification). Based on Hf-W isotopic systematics, this assumption may not be correct. For example, Nimmo et al. (2010) estimated that 30-80% of impactor cores must have equilibrated whereas Rudge et al. (2011) estimated the equilibrated fraction to be 36% (see also Rubie et al., 2011). Here we repeat the least squares refinements for simulation 4:1-0.25-7, based on the composition-distance model of Fig. 6, with fractions of equilibrated metal that vary from 1.0 to 0.0. The results (Fig. 9a) show that the goodness of fit becomes progressively worse as the fraction of equilibrated metal ($\xi$) decreases. Excellent fits are obtained with $\xi \geq 0.7$ but with $\xi < 0.6$ the value of $\chi_{red}^2$ increases significantly to >13. The reason for this trend is that, with less equilibration, the Ni and Co mantle concentrations become increasingly dominated by a low pressure signature – i.e. too much Ni and Co is transported into the Earth's core in metal that has only equilibrated at low pressure.

We have not considered the effect of impactor size. Incomplete equilibration is more likely in the case of large impactors (embryos) than small bodies (planetesimals). This issue will be addressed in future studies.

*7.3 Extent of silicate equilibration*

According to the model of Deguen et al. (2011), only a very limited fraction of a target's silicate mantle equilibrates with the metal of an impactor's core (Eq. 9, Fig. 3). This fraction depends on the radius of the impacting core and, in the case of the simulations discussed here lie in the range 0.00064 to 0.014 for planetesimal cores and 0.008 to 0.11 for embryo cores. Here we consider the possibility that the model of Fig. 3 underestimates (or even overestimates) the fraction of equilibrating mantle. For example, at the base of the hemispherical melt pool, the descending plume could spread out sideways and entrain additional silicate magma that would also be involved in the equilibration process (Fig. 3; Deguen et al., 2014). In addition, the model of Fig. 3 applies to high-angle impacts and the situation should be different for low-angle impacts.

Figure 9b shows the effect on the goodness-of-fit for simulation 4:1-0.25-7 when the fraction of silicate mantle that equilibrates, as determined by Eq. 9, is multiplied by a factor Φ that we vary from 0.5 to 5.0. The results show that the model of Deguen et al. (2011) provides an excellent estimate; when Φ deviates significantly from unity, the goodness-of-fit deteriorates rapidly. However, Φ and the equilibrated fraction of impactor cores are at least weakly correlated (dashed line in Fig. 9b); with only 80% of metal equilibrating, values of Φ ≥ 2 cannot be excluded.

*7.4 Equilibration temperature*

Temperatures of metal-silicate equilibration are determined as a function of equilibration pressure using Eq. 8, which is a model for temperatures lying approximately midway between the peridotite liquidus and solidus. As described above (section 2.4) there are considerable uncertainties in peridotite melting temperatures because of discrepancies between different studies (at least ±100 K even in low pressure studies). We therefore examine the effect of adjusting the temperatures of Eq. 8 by a factor $\Delta T$ on the goodness-of-fit for simulation 4:1-0.25-7 based on the composition-distance model of Fig. 6b. The results are shown in Fig. 9c for $\Delta T$ values ranging from -200 K to +200 K. This shows that the best fits are obtained with $\Delta T$ in the range -50 to +30 K, thus indicating that Eq. 8 provides a reliable estimate of metal-silicate equilibration temperatures. Equilibration pressures and temperatures are correlated so that as $\Delta T$ increases, equilibration pressures, as given by $f_P$, decrease (Fig. 9c)

## 8. Uncertainties, caveats and problems for the future

The differentiation model described above is based on six Grand Tack N-body simulations in which it is assumed that all collisions are perfect, i.e. all the mass in each pairwise collision is conserved in the mass of the final body. This assumption greatly simplifies the N-body calculation but could have consequences for the differentiation model. These consequences can be divided into two different categories: (a) imperfect giant impacts and (b) planetesimal collisional grinding. (a) Giant hit'n'run impacts and other debris-generating impacts can occur in the protoplanetary disk (Agnor & Asphaug, 2004). Dynamically, a consequence of this debris generation is a slight prolonging of the length of the giant impact phase of planet formation and planets on more circular orbits, but the final planets appear to be drawn from the same distribution in terms of mass and semi-major axis (Kokubo and Genda, 2010, Chambers, 2013). Debris generated from imperfect accretion tends to be re-accreted by the target body that generated it, so mass is ultimately conserved. Chambers (2013) found that this process did not lead to the systematic removal of mantle material, so this process by itself is unlikely to play a significant role on core formation and planetary differentiation. This conclusion is also supported by determinations in this study of core mass fractions for the Earth that are almost identical to the observed value (Table 5)

(b) Collisional grinding of the planetesimal population may have a larger effect since it may bias the compositional ingredients delivered during planet formation. Collisional grinding is the process of planetesimal on planetesimal impacts, similar to asteroid family creation today in the Main Asteroid Belt, that eventually comminutes small (~<1000 km) planetesimals to dust. Dynamically, this process removes small bodies from the protoplanetary disk and so decreases the amount of dynamical friction in the disk. Regarding planetary differentiation and core formation, if the planetesimals all have the same chondritic composition, this process does not affect the final chemical composition of the planets. However, if collisional grinding involves achondritic debris generated from imperfect giant impacts, then this process could lead to an unaccounted for chemical change in the disk. This is a topic for future study.

An estimate of the uncertainties on the results of this study can be obtained by examining the ranges of fitting parameter values obtained in the six N-body simulations (Table 4). For example, metal-silicate equilibration pressures are predicted to lie in the range $0.58 \times P_{CMB}$ to $0.72 \times P_{CMB}$, where $P_{CMB}$ is the core-mantle boundary pressure in the accreting bodies at the time of impact. Uncertainties on the composition-distance model for primitive bodies can also be evaluated from Fig. 6 and Table 4. A more rigorous analysis of uncertainties and an investigation of possible correlations between fitted parameters will provided by Jacobson et al. (in preparation).

Additional uncertainties arise from the metal-silicate partitioning models. First, it is necessary to make very large extrapolations of the partitioning data (e.g. extrapolations from 20-25 GPa and 2500-3000 K to 80-100 GPa and 4000 K). The error bars in Figs. 5 and 7 show the uncertainties on the final Earth mantle compositions based on error propagation. In future, high-quality experimental partitioning data are required for all elements of interest to conditions of 100 GPa and >4000 K (Fischer et al., 2014b). However, there are significant technical problems that can affect the results of such experiments (see Supplementary Data). Second, we have used one specific set of partitioning models (Table S2, Supplementary data), whereas for most of the elements considered here there have been a significant number of experimental studies, each of which has proposed a different partitioning model (e.g. Table 3 in Rubie et al., 2007 and 2014b). The effect of using alternative $K_D$ models for Ni and Co is that fitted parameter values change slightly and the quality of the fits can be affected (see Supplementary Data). The obvious solution to this problem is to fit $K_D$ to all available experimental data for each element (e.g. Righter, 2011). However, this has to be done extremely carefully because of differences in experimental techniques used in the various studies. In addition, rigorous thermodynamic models need to be fit to the data and simple parameterisations (e.g. Righter, 2011) may be unreliable when extrapolating.

Obvious goals for the future include incorporating volatile elements in the model. By including sulphur, the behaviour of W and Mo during core formation can also be examined (the partitioning of these elements is significantly affected by the S content of liquid Fe - see Wade et al., 2012). For some elements, the effects of both metal and silicate melt compositions on partitioning need to be taken into account (e.g. Wade et al., 2012).

Finally, the incorporation of Hf-W isotopic evolution into the combined accretion/differentiation model described here will help to further distinguish between successful and unsuccessful models. In particular, by modelling the Earth's tungsten anomaly it may be possible to better constrain the extent of metal equilibration during core formation (Rudge et al., 2010; Nimmo et al., 2010).

## 9. Summary and conclusions

We have successfully integrated a multistage core-mantle differentiation model with Grand Tack N-body accretion simulations. The combined model enables the compositional evolution of the mantles and cores of the Earth and other terrestrial planets to be modelled simultaneously. In the six simulations studied here, an Earth-like planet can be generated with mantle concentrations of non-volatile elements and $H_2O$ that are close to the concentrations estimated for the Earth's primitive mantle (Fig. 7, Table 5). The results predict that the Earth's core contains 2-4 wt% oxygen and 8-9 wt% silicon. The exact compositional paths followed by the mantle and core of the Earth during accretion are model dependent due to the stochastic nature of the accretion process (Fig. 7). For example, water is accreted to the Earth mainly after 60-80% of the planet has accreted. However, in one simulation (4:1-0.5-8), there is a large increase in $H_2O$ content at ~33% accretion due to collisions with a series of $H_2O$-bearing planetesimals (Fig. 7a).

The model places strong constraints on the oxidation state of primitive embryos and planetesimals as a function of their heliocentric distances of origin in the early Solar System. Acceptable results are obtained only when bodies that formed close to the Sun (<0.9-1.2 AU) were highly reduced. At greater distances, primitive bodies were increasingly oxidised and were fully oxidised with high concentrations of $H_2O$ beyond 6-7 AU. This oxidation gradient may be explained by the net inward flow of ice-covered dust from beyond the snow line in the solar nebula, with sublimed $H_2O$ causing oxidization of Fe, although details of this mechanism are still disputed. However, regardless of its cause, the variation in oxidation states is consistent with those recorded by meteorites (e.g. from enstatite chondrites to CI chondrites). Other composition-distance models provide unacceptably poor results. For example, models in which primitive body compositions are constant with heliocentric distance and those in which compositions from close to the Sun are oxidized and compositions from greater distances are reduced can be excluded. In addition, models of core formation under oxidizing conditions (Rubie et al., 2004; Siebert et al., 2013) fail badly when mass balance is considered.

The Mars analogues that form in the simulations are very variable in terms of both oxidation state (e.g. mantle FeO content) and core mass fraction. Considering only Mars analogues that accrete within the first 10 Myr (Dauphas and Pourmand, 2011), the oxidation state depends critically on the heliocentric distance at which the Mars-forming embryo originates (Fig. 6). Simulation i-4:1-0.8-6 produces the most realistic Mars analogue in terms of mantle composition and core mass fraction and originates from an embryo that formed at 1.58 AU (Table 7). Results of this simulation suggest that the Martian core contains negligible concentrations of the light elements Si and O.

Finally, we make predictions about the compositions of the core and mantle of Venus (Table 6). In general, the predicted compositions are close to those of the Earth, although in two simulations the mantle FeO content is relatively low (4-6 wt%).

**Acknowledgements**

D.C.R., S.A.J., A.M., J.deV., and D.P.O. were supported by the European Research Council (ERC) Advanced Grant "ACCRETE" (contract number 290568). D.P.O. was also supported by grant NNX09AE36G from NASA's Planetary Geology and Geophysics research program. F.N. acknowledges support by NASA Origins NNX11AK60G. We thank John Hernlund for discussions and two anonymous reviewers for constructive and helpful criticisms.

**Appendix A. Supplementary material**

Supplementary Data associated with this article can be found in the online version at…..
The complete set of chemical evolution results for the mantles and cores of planets in simulation i-4:1-0.8-6 can be found in Table S3 of the Supplementary Data. Results of the other simulations are available from the corresponding author upon request.

**References**

Abe, Y. (1993) Physical state of the very early Earth. Lithos 30, 223-235.

Alfè, D., Gillan, M.J., Price, G.D. (2002) Composition and temperature of the Earth's core constrained by combining ab initio calculations and seismic data. Earth Planet. Sci. Lett. 195, 91-98.

Alfè, D.,Gillan, M.J., Price, G.D. (2007) Temperature and composition of the Earth's core. Contemporary Physics 48, 63-80.

Allègre, C.J., Poirier, J.-P., Humler, E., Hofmann, A.W. (1995) The chemical composition of the Earth. Earth Planet. Sci. Lett. 134, 515-526.

Andrault, D., Bolfan-Casanova, N., Lo Nigro, G., Bouhifd, A., Garbarino, G., Mezouar, M. (2010) Solidus and liquidus profiles of chondritic mantle: Implication for melting of the Earth across its history. Earth Planet. Sci. Lett. 304, 251-259.

Armitage, P.J. (2011) Dynamics of protoplanetary disks. Annu. Rev. Astron. Astrophys. 49, 195-236.

Agnor, C., Asphaug, E. (2004) Accretion efficiency during planetary collisions. Astrophysical Journal Letters 613, L157-L160.

Badro, J., Fiquet, G., Guyot, F., Gregoryanz, E., Occelli, F., Antonangeli, D., d'Astotu, M. (2007) Effect of light elements on the sound velocities in solid iron: Implications for the composition of Earth's core. Earth Planet. Sci. Lett. 254, 233-238.

Benz, W., Cameron, A.G.W., Melosh, H.J. (1989) The origin of the Moon and the single impact hypothesis. III. Icarus 81, 113-131.

Bond, J.C., Lauretta, D.S., O'Brien, D.P. (2010) Making the Earth: Combining dynamics and chemistry in the Solar System. Icarus 205, 321-337, doi:10.1016/j.icarus.2009.07.037.

Chambers, J.E. (2013) Late-stage planetary accretion including hit-and-run collisions and fragmentation. Icarus 224, 43-56.

Ćuk, M., Stewart, S.T. (2012) Making the Moon from a fast-spinning Earth: A giant impact followed by resonant despinning. Science 338, 1047-1052.

Cuzzi, J.N., Zahnle, K.J. (2004) Material enhancement in protoplanetary nebulae by particle drift through evaporation fronts. The Astrophysical Journal 614, 490-496.

Dahl, T. W., and D. J. Stevenson (2010) Turbulent mixing of metal and silicate during planet accretion-and interpretation of the Hf-W chronometer, Earth and Planetary Science Letters, 295, 177-186.

Dauphas, N., Pourmand, A. (2011) Hf-W-Th evidence for rapid growth of Mars and its status as a planetary embryo. Nature 473, 489-492, doi:10.1038/nature10077.

Deguen, R., Olson, P., Cardin, P. (2011) Experiments on turbulent metal- silicate mixing in a magma ocean, Earth and Planetary Science Letters, 310, 303-313.

Deguen, R., Landeau, M., Olson, P. (2014) Turbulent metal-silicate mixing, fragmentation, and equilibration in magma oceans. Earth Planet. Sci. Lett. 391, 274-287.

de Vries, J., Nimmo, F., Melosh, H.J., Jacobson, S., Morbidelli, A., Rubie, D.C. (2014) Melting due to impacts on growing proto-planets. 45th Lunar and Planetary Science Conference, Abstract #1896.

Drake, M.J., Righter, K. (2002) Determining the composition of the Earth. Nature 416, 39-44.

Dreibus, G., Wänke, H. (1985) Mars, a volatile-rich planet. Meteoritics 20, 367-381.

Dreibus, G., Wänke, H. (1987) Volatiles on Earth and Mars: A comparison. Icarus 71, 225-240.

Duncan, M. J., Levison, H. F., Lee, M. H. (1998) A multiple time step symplectic algorithm for integrating close encounters. Astron. J. 116, 2067-2077.

Elser S., Meyer M.R., Moore B. (2012) On the origin of elemental abundances in the terrestrial planets. Icarus 221, 859-874.

Fiquet, G., Auzende, A.L., Siebert, J., Corgne, A., Bureau, H., Ozawa, H., Garbarino, G. (2010) Melting of peridotite to 140 Gigapascals. Science 329, 1516-1518.

Fischer, R.A., Campbell, A.J., Shofner, G.A., Lord, O.T., Dera, P., Prakapenka, V.B. (2011) Equation of state and phase diagram of FeO. Earth Planet. Sci. Lett. 304, 496-502.

Fischer, R.A., Campbell, A.J., Caracas, R., Reaman, D.M., Heinz, D.L., Dera, P., Prakapenka, V.B. (2014a) Equations of state in the Fe-FeSi system at high pressures and temperatures, J. Geophys. Res. Solid Earth, 119, 2810–2827, doi:10.1002/2013JB010898.

Fischer, R.A., Campbell, A.J., Rubie, D.C., Frost, D.J., Miyajima, N., Pollack, K, Harries, D. (2014b) Experimental constraints on the core's Si and O contents from equations of state and metal-silicate partitioning. Abstract, Goldschmidt 2014.

Frost D.J., Mann U., Asahara Y., Rubie D.C. (2008) The redox state of the mantle during and just after core formation. Phil. Trans. R. Soc. A 366, 4315-4337. doi: 10.1098/rsta.2008.0147.

Frost, D.J., Asahara, Y., Rubie, D.C., Miyajima, N., Dubrovinsky, L.S., Holzapfel, C., Ohtani, E., Miyahara, M., Sakai, T. (2010) The partitioning of oxygen between the Earth's mantle and core. Journal of Geophysical Research 115, B02202, doi:10.1029/2009JB006302.

Gessmann, C.K., Rubie, D.C., McCammon, C. (1999) Oxygen fugacity dependence of Ni, Co, Mn, Cr, V and Si partitioning between liquid metal and magnesiowüstite at 9-18 GPa and 2200°C. Geochim. Cosmochim. Acta 63, 1853-1863.

Greenberg, R., Hartmann, W.K., Chapman, C.R., Wacker, J.F. (1978) Planetesimals to planets—numerical simulation of collisional evolution. Icarus 35. 1-26

Grossman L., Fedkin A. V. and Simon S. B. (2012) Formation of the first oxidized iron in the solar system. Meteoritics & Planetary Science 47, 2160-2169.

Hansen, B.M.S. (2009) Formation of the Terrestrial Planets from a Narrow Annulus. ApJ 703, 1131-1140.

Herzberg, C., Zhang, J., 1996. Melting experiments on anhydrous peridotite KLB-1: compositions of magmas in the upper mantle and transition zone. J. Geophy. Res. 101, 8271-8295.

Hirschmann, M.M. (2006) Water, melting, and the deep Earth $H_2O$ cycle. Annu. Rev. Earth Planet. Sci. 34, 629-653.

Hsieh, H.H., Jewitt, D. (2006) A population of comets in the main asteroid belt. Science 312, 561-563.

Huang, H., Fei, Y., Cai, L., Jing, F., Hu, X., Xie, H., Zhang, L., Gong, Z. (2012) Evidence for an oxygen-depleted liquid outer core of the Earth. Nature 479, 513-516.

Jacobson, S.A., Morbidelli, A. (2014) Lunar and terrestrial planet formation in the Grand Tack scenario. Proc. Royal Soc. A, submitted.

Jacobson, S.A., Morbidelli, A., Raymond, S.N., O'Brien, D.P., Walsh, K.J., Rubie, D. C. (2014) Highly siderophile elements in the Earth's mantle as a clock for the Moon-forming impact. Nature 508, 84-87.

Javoy, M. (1995) The integral enstatite chondrite model of the Earth. Geophys. Res. Lett. 22, 2219-2222.

Jewitt, D., Yang, B., Haghighipour, N. (2009) Main-belt comet P/2008 R1 (Garradd). The Astronomical Journal 137, 4313-4321.

Kegler, P., Holzheid, A., Frost, D.J., Rubie, D.C., Dohmen, R., Palme, H. (2008) New Ni and Co metal-silicate partitioning data and their relevance for an early terrestrial magma ocean. Earth Planet. Sci. Lett. 268, 28-40.

Keil, K. (1968) Mineralogical and chemical relationships among enstatite chondrites. Journal of Geophysical Research 73, 6945-6976.

Kerridge, J.F., Bunch, T.E. (1979) Aqueous activity. In: Gehrels, T. (Ed.), Asteroids. University of Arizona Press, Tucson, AZ, pp. 745-764.

Kokubo, E., Ida, S. (1998) Oligarchic growth of protoplanets. Icarus 131, 171-78.

Kokubo, E, Genda, H. (2010) Formation of terrestrial planets from protoplanets under a realistic accretion condition. Astrophysical Journal Letters 714, L21-L25.

Krot, A.N., Petaev, M.I., Scott, E.R.D., Choi, B.G., Zolenski, M.E., Keil, K. (1998) Progressive alteration in CV3 chondrites: more evidence for asteroidal alteration. Meteoritics Planet. Sci. 33, 1065-1085.

Küppers, M., O'Rourke, L., Bockelée-Morvan, D., Zakharov, V., Lee, S., von Allmen, P., Carry, B., Teyssier, D., Marston, A., Müller, T., Crovisier, J., Barucci, M.A., Moreno, R. (2014) Localized sources of water vapour on the dwarf planet (1) Ceres. Nature 505, 525-527.

Levison, H.F., Morbidelli, A., Tsiganis, K., Nesvorny, D. Gomes, R.S, (2011) Late orbital instabilities in the outer planets induced by interaction with a self-gravitating planetesimal disk. Astron. J. 142, 152.

Liebske, C., Frost, D.J. (2012) Melting phase relations in the MgO-MgSiO$_3$ system between 16 and 26 GPa: Implications for melting in Earth's deep interior. Earth and Planetary Science Letters, 345-348, 159-170.

Krot A. N., Fegley B. and Lodders K. (2000). Meteoritical and astrophysical constraints on the oxidation state of the Solar Nebula. In: Protostars and Planets IV. Mannings, V., Boss, A.P., Russell, S.S. (Eds), University of Arizona Press**:** 1019-1054.

Mann, U., Frost, D.J., Rubie, D.C. (2009) Evidence for high-pressure core-mantle differentiation from the metal-silicate partitioning of lithophile and weakly siderophile elements. Geochim. Cosmochim. Acta 73, 7360-7386, doi:10.1016/j.gca.2009.08.006.

Masset, F., Snellgrove, M. (2001) Reversing type II migration: resonance trapping of a lighter giant protoplanet. Mon. Not. R. Astron. Soc. 320, L55–L59.

McDonough, W.F. (2003) Compositional model for the Earth's core, in: Carlson, R.W. (Ed.), Treatise on Geochemistry, Volume 2-The Mantle and Core. Elsevier-Pergamon, Oxford, pp. 547-568.

McSween, H.Y. (2003) Mars. In: Davis, A.M. (Ed.), Treatise on Geochemistry, Volume 1-Meteorites, Comets and Planets. Elsevier-Pergamon, Oxford, pp. 601-621.

McSween H.Y., Ghosh A., Grimm R.E., Wilson L., and Young E.D. (2003) Thermal Evolution Models of Asteroids.  In Bottke W.F. Jr. et al. (eds) Asteroids III.  University of Arizona Press, Tucson, 559-572.

Morbidelli, A., Chambers, J., Lunine, J.I., Petit, J.M., Robert, F., Valsecchi, G.B., Cyr, K.E., 2000. Source regions and timescales for the delivery of water to Earth. Meteoritics and Planetary Science, 35, 1309-1320.

Morbidelli, A., Crida, A. (2007) The dynamics of Jupiter and Saturn in the gaseous protoplanetary disk. Icarus 191, 158–171.

Morbidelli, A., Tsiganis, K., Crida, A., Levison, H.F., Gomes, R.S., (2007) Dynamics of the giant planets of the Solar System in the gaseous protoplanetary disk and their relationship to the current orbital architecture. Astron. J. 134, 1790–1798.

Morbidelli, A., Lunine, J. I., O'Brien, D. P., Raymond, S. N., Walsh, K. J., (2012) Building terrestrial planets. Annu. Rev. Earth Planet. Sci. 40, 251–275.

Morishima, R., Golabek, G.J., Samuel, H. (2013) N-body simulations of oligarchic growth of Mars: Implications for Hf-W chronology. Earth Planet. Sci. Lett. 366, 6-16.

Münker, C., Pfänder, J.A., Weyer, S., Büchl, A., Kleine, T., Mezger, K., 2003. Evolution of planetary cores and the Earth-Moon system from Nb/Ta systematics. Science 301, 84-87.

Nelder, J.A., Mead, R., 1965. A simplex method for function minimization. Computer Journal 7, 308-313.

Nimmo, F., Agnor, C.B., 2006. Isotopic outcomes of N-body accretion simulations: Constraints on equilibration processes during large impacts from Hf/W observations. Earth Planet. Sci. Lett. 243, 26-43.

Nimmo, F., O'Brien, D.P., Kleine, T. (2010), Tungsten isotopic evolution during late-stage accretion: constraints on Earth-Moon equilibration, Earth Planet. Sci. Lett. 292, 363-370.

Nomura, R., Hirose, K., Uesugi, K., Ohishi, Y., Tsuchiyama, A., Miyake, A., Ueno, Y. (2014) Lower core-mantle boundary temperature inferred from the solidus of pyrolite. Science 343, 522-525.

O'Brien, D.P., Morbidelli, A., Levison, H.F. (2006) Terrestrial planet formation with strong dynamical friction. Icarus 184, 39-58.

O'Brien, D.P., Walsh, K.J., Morbidelli, A., Raymond, S.N., Mandell, A.M. (2014) Water delivery and giant impacts in the 'Grand Tack' scenario. Icarus, submitted.

Okuchi, T. (1997) Hydrogen partitioning into molten iron at high pressure: Implications for Earth's core. Science 278, 1781-1784.

O'Neill, H.St.C., Palme, H. (1998) Composition of the silicate Earth: Implications for accretion and core formation. In: Jackson, I. (ed.) The Earth's Mantle: Structure, Composition and Evolution – The Ringwood Volume, pp. 3-126. Cambridge University Press, Cambridge.

Palguta, J., Schubert, G., Travis, B.J. (2010) Fluid flow and chemical alteration in carbonaceous chondrite parent bodies. Earth and Planetary Science Letters 296, 235-243.

Palme, H., O'Neill, H.St.C. (2013) Cosmochemical estimates of mantle composition, in: Carlson, R.W. (Ed.), Treatise on Geochemistry 2$^{nd}$ edition, Volume 2-The Mantle and Core. Elsevier-Pergamon, Oxford, pp. 1-39.

Pierens, A., Nelson, R. P. (2008) Constraints on resonant-trapping for two planets embedded in a protoplanetary disc. Astron. Astrophys. 482, 333–340.

Press, WH, Teulolsky, SA, Vetterling, WT, Flannery, B.P. (2002) Numerical Recipes in C++, 2$^{nd}$ edition. Cambridge Univ. Press, Cambridge, UK, pp. 413-417.

Ricolleau, A., Fei, Y., Corgne, A., Siebert, J., Badro, J. (2011) Oxygen and silicon contents of Earth's core from high pressure metal-silicate partitioning experiments. Earth Planet. Sci. Lett. 310, 409-421.

Righter, K. (2011) Prediction of metal–silicate partition coefficients for siderophile elements: An update and assessment of PT conditions for metal–silicate equilibrium during accretion of the Earth. Earth and Planetary Science Letters 304, 158-167.

Righter, K., Chabot, N.L. (2011) Moderately and slightly siderophile element constraints on the depth and extent in melting in early Mars. Meteoritics and Planetary Science 46, 157-176.

Ringwood, A.E. (1984) The Earth's core: Its composition, formation and bearing on the origin of the Earth. Proc. Royal Soc. Series A 395, 1-46.

Robinson, M.S., Taylor, G.J. (2001) Ferrous oxide in Mercury's crust and mantle. Meteoritics and Planetary Science 36, 841-847.

Rubie, D.C. Melosh, H.J., Reid, J.E., Liebske, C., Righter, K. (2003). Mechanisms of metal-silicate equilibration in the terrestrial magma ocean. Earth Planet. Sci. Lett. 205, 239-255.

Rubie, D.C., Gessmann, C.K., Frost, D.J., 2004. Partitioning of oxygen during core formation on the Earth and Mars. Nature 429, 58-61.

Rubie, D.C., Nimmo, F., Melosh, H.J. (2007) Formation of the Earth's core, in: Stevenson, D. (Ed.), Treatise on Geophysics, Volume 9-Evolution of the Earth. Elsevier, Amsterdam, pp. 51-90.

Rubie, D.C., Frost, D.J., Mann, U., Asahara, Y., Tsuno, K., Nimmo, F., Kegler, P., Holzheid, A., Palme, H. (2011) Heterogeneous accretion, composition and core-mantle differentiation of the Earth. Earth and Planetary Science Letters 301, 31-42, doi: 10.1016/j.epsl.2010.11.030

Rubie, D.C., Jacobson, S.A., Morbidelli, A., O'Brien, D.P., Young, E.D. (2014a) Accretion and differentiation of the terrestrial planets: Implications for the compositions of early-formed Solar system bodies. 45[th] Lunar and Planetary Science Conference, Abstract #1734.

Rubie, D.C., Nimmo, F., Melosh, H.J. (2014b) Formation of the Earth's core, in: Stevenson, D. (Ed.), Treatise on Geophysics 2[nd] edition, Volume 9-Evolution of the Earth. Elsevier, Amsterdam, in press.

Rudge, J.F., Kleine, T., Bourdon, B., 2010. Broad bounds on Earth's accretion and core formation constrained by geochemical models. Nature Geoscience 3, 439-443, DOI: 10.1038/NGEO872

Samuel, H. (2012), A re-evaluation of metal diapir breakup and equilibration in terrestrial magma oceans, Earth and Planetary Science Letters, 313-314, 105-114.

Sasaki, T., Abe, Y. (2007) Rayleigh-Taylor instability after giant impacts: Imperfect equilibration of the Hf-W system and its effect on core formation age. Earth Planets Space 59, 1035-1045.

Schönbächler, M., Carlson, R.W., Horan, M.F., Mock, T.D., Hauri, E.H., 2010. Heterogeneous accretion and the moderately volatile element budget of Earth. Science 328, 884-887, DOI: 10.1126/science.1186239.

Sharp, Z.D., McCubbin, F.M., Shearer, C.K. (2013) A hydrogen-based oxidation mechanism relevant to planetary formation. Earth Planet. Sci. Lett. 380, 88-97.

Siebert J, Badro J, Antonangeli D, Ryerson FJ (2012) Metal–silicate partitioning of Ni and Co in a deep magma ocean. Earth and Planetary Science Letters 321-322 (2012) 189–197.

Siebert, J., Badro, J., Antonangeli, D., Ryerson, F.J. (2013) Terrestrial accretion under oxidizing conditions. Science 339, 1194-1197.

Sohl, F., Spohn, T. (1997) The interior structure of Mars: Implications from SNC meteorites. J. Geophys. Res. 102, 1613-1635.

Taylor, G.J. (2013) The bulk composition of Mars. Chemie der Erde 73, 401-420.

Taylor, G.J., Scott, E.R.D. (2003) Mercury, in: Davis, A.M. (Ed.), Treatise on Geochemistry, Volume 1-Meteorites, Comets and Planets. Elsevier-Pergamon, Oxford, pp. 477-485.

Tonks, W.B., Melosh, H.J. (1993) Magma ocean formation due to giant impacts, J. Geophys. Res. 98, 5319-5333.

Treiman, A.H. (2009) Venus' bulk and mantle compositions: Are Venus and Earth really twins? 40[th] Lunar and Planetary Science Conference, abstract #2016.

Trønnes, R.G., Frost, D.J. (2002) Peridodite melting and mineral-melt partitioning of major and minor elements at 22-24.5 GPa. Earth Planet. Sci. Lett. 197, 117-131.

Tsuno, K., Frost, D.J., Rubie, D.C. (2013) Simultaneous partitioning of silicon and oxygen into the Earth's core during early Earth differentiation. Geophysical Research Letters 40, 66–71, doi:10.1029/2012GL054116

Turcotte D. L., and Schubert G. (2002) Geodynamics, Cambridge Univ. Press, Cambridge, 456pp.

Tuff J, Wood, BJ, Wade J (2011) The effect of Si on metal–silicate partitioning of siderophile elements and implications for the conditions of core formation. Geochimica et Cosmochimica Acta 75, 673-690.

Wade , J., Wood, B.J. (2005) Core formation and the oxidation state of the Earth. Earth Planet. Sci. Lett. 236, 78-95.

Wade, J, Wood BJ, Tuff J (2012) Metal–silicate partitioning of Mo and W at high pressures and temperatures: Evidence for late accretion of sulphur to the Earth. Geochimica et Cosmochimica Acta 85, 58–74

Walsh KJ, Morbidelli A, Raymond SN, O'Brien DP, Mandell AM (2011) A low mass for Mars from Jupiter's early gas-driven migration, Nature 475, 206-209.

Wood B. J., Walter, M. J., and Wade J. (2006) Accretion of the Earth and segregation of its core. Nature 441, 825-833, doi:10.1038/nature04763.

Wood, J.A. (2000) Pressure and temperature profiles in the solar nebula. Space Science Rev. 92, 87-93.

Young, E.D., Ash, R.D., England, P., Rumble III, D. (1999) Fluid flow in chondritic parent bodies: Deciphering the compositions of planetesimals. Science 286, 1331-1335.

Young, E.D. ( 2001) The hydrology of carbonaceous chondrites parent bodies and the evolution of planet progenitors. Phil. Trans. R. Soc. Lond. A 359, 2095-2110.

Zhang, K., Pontoppidan, K.M., Salyk, C., Blake, G.A. (2013) Evidence for a snow line beyond the transitional radius in the TW Hya protoplanetary disk. The Astrophysical Journal 766:82 (12pp).

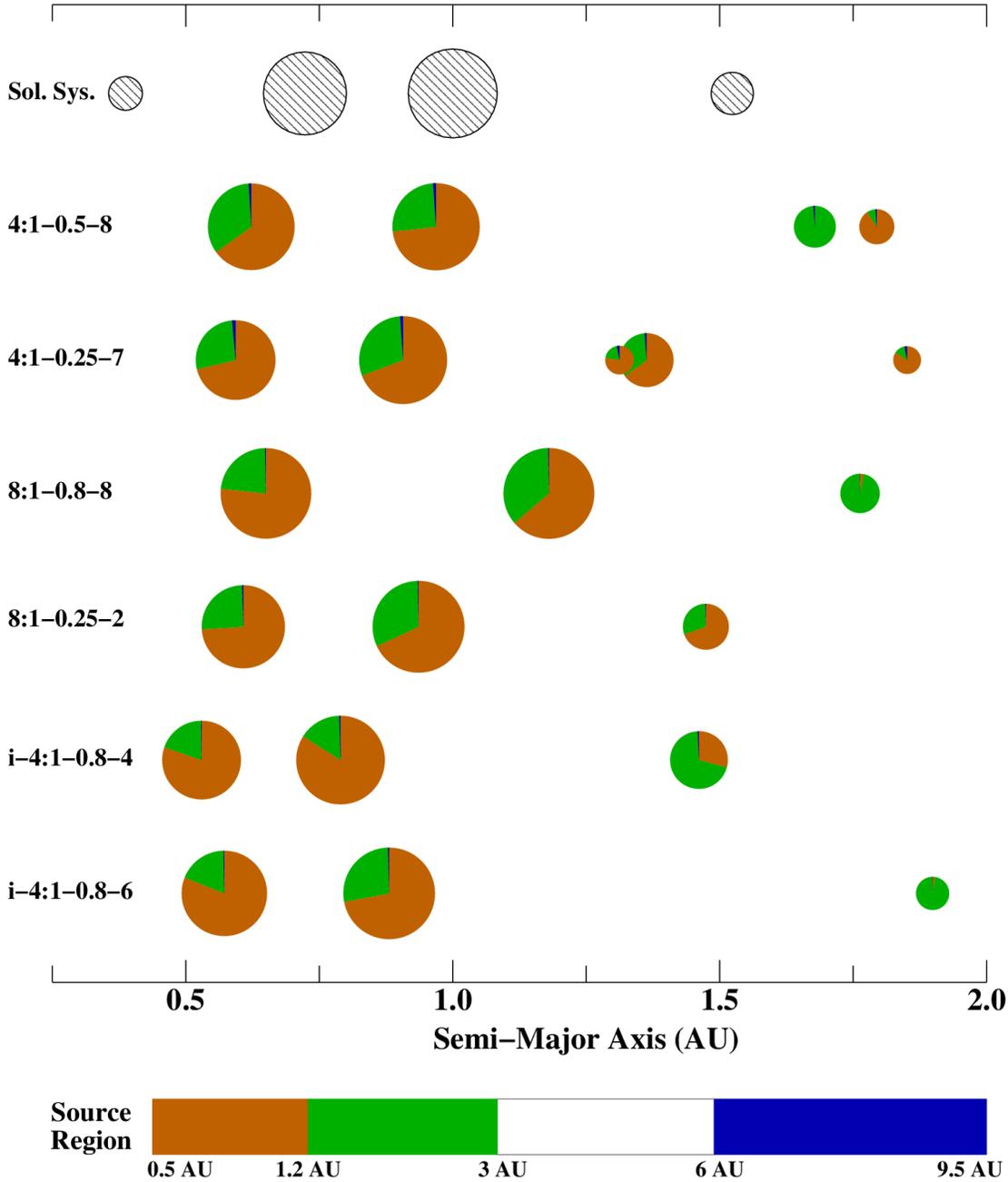

**Figure 1**. Mass and location of the final planets in the six Grand Tack simulations of this study. The actual planets of the Solar System are shown at the top for comparison. The colored segments show the proportions of accreted material that originates from 0.5-1.2 AU (brown), 1.2-3 AU (green) and 6-9.5 AU (blue), respectively. Note that no material originates between 3 and 6 AU because the formation of Jupiter and Saturn cleared all bodies from this region.

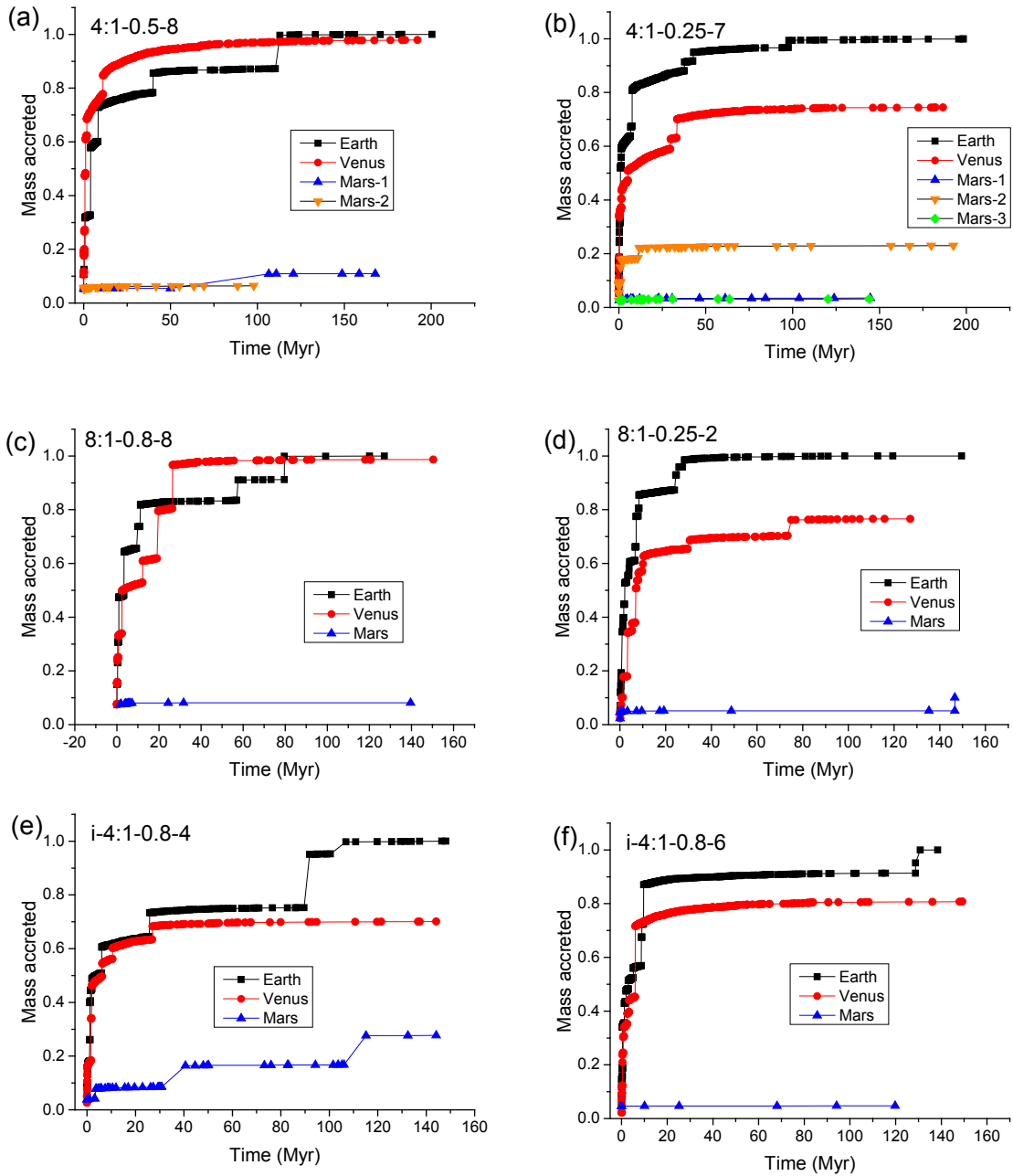

**Figure 2**. Accretion histories of the model planets in the six Grand Tack simulations. Mass accreted, as a fraction of one Earth mass, is plotted against time. Each symbol shows the mass immediately following an impact event. Note that the lines connecting the symbols do not accurately represent the mass evolution path which in reality involves a series of vertical steps.

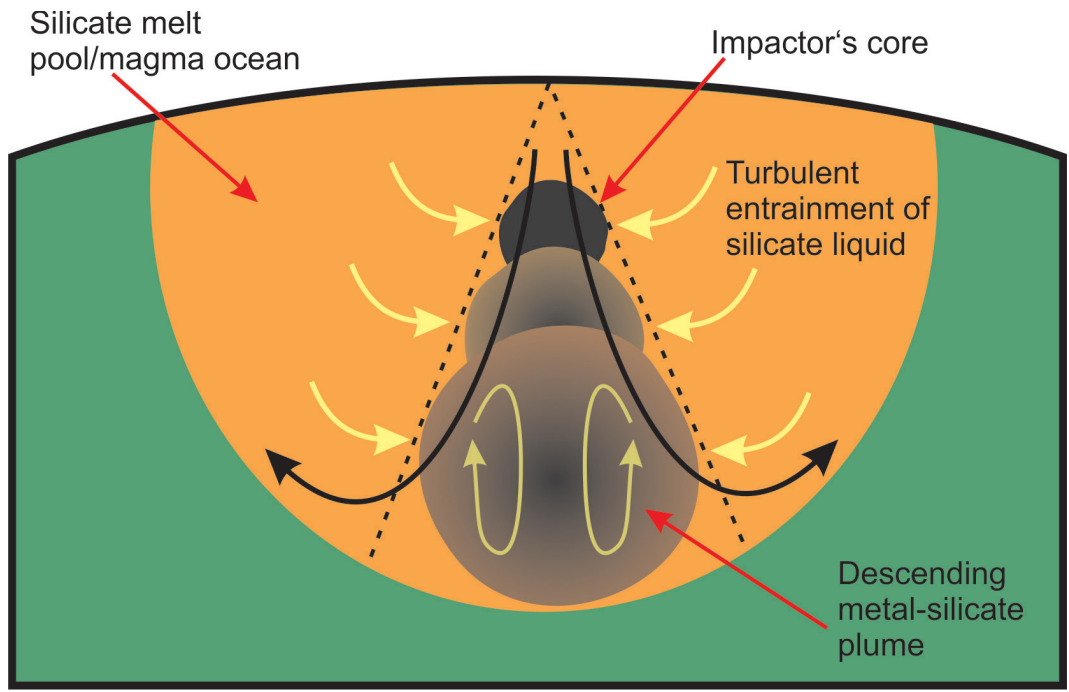

**Figure 3**. Descent of an impactor's metallic core through silicate liquid in an impact-generated melt pool/magma ocean. The descending core turbulently entrains silicate liquid in a plume-like structure that expands with increasing depth. The volume fraction $\phi$ of metal in the metal-silicate plume is calculated from Eq. 9 and enables the mass fraction of the proto-planet's mantle that equilibrates with the metal to be determined. (After Deguen et al., 2011.)

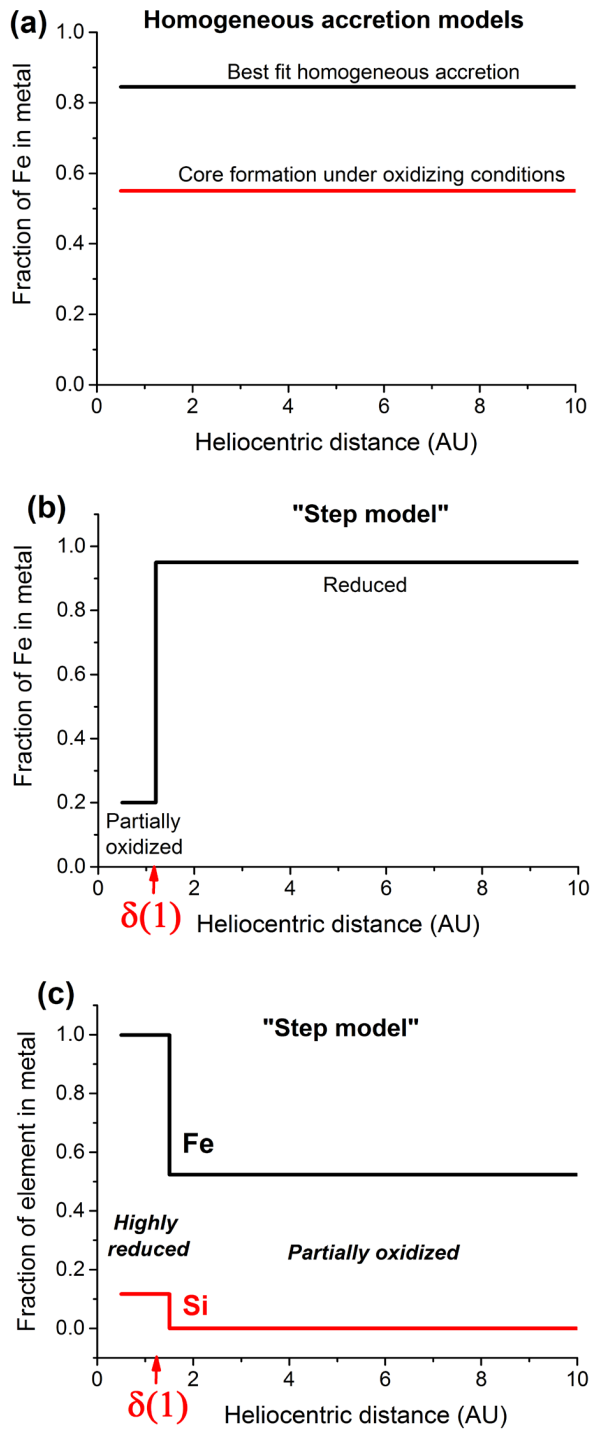

**Figure 4**. Simple composition-distance models for primitive embryos and planetesimals in the early solar system that are tested here. (a) Two models that result in homogeneous accretion. The "best-fit" model results from the best fit value of $X_{Fe}^{met}$ = 0.84-0.85. For the "core formation under oxidizing conditions" model, $X_{Fe}^{met}$ is fixed at 0.55 in order to test this model (see text). (b) "Step model" in which compositions close to the Sun are oxidized and those further out are reduced. (c) "Step model" in which bodies that originate close to the Sun are highly reduced and those from beyond distance δ(1) are partially oxidized.

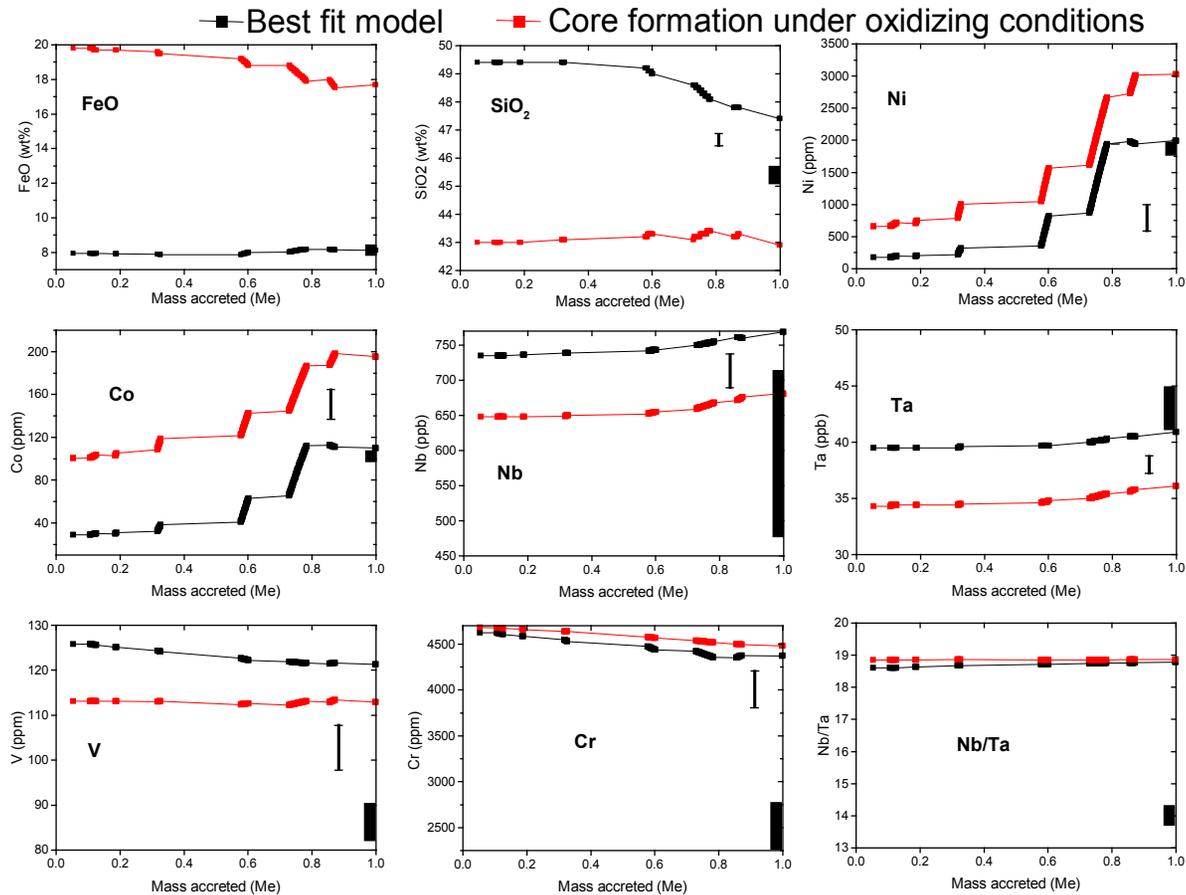

**Figure 5**. Evolution of element concentrations in the Earth's mantle as a function of mass accreted for two homogeneous accretion/core formation models for Grand Tack simulation 4:1-0.5-8 (Fig. 4a, Table 2). Estimated primitive mantle concentrations (Palme and O'Neill, 2013) are shown as a black bar on the right-hand axis of each graph. The error bars indicate the magnitude of propagated uncertainties on the final concentrations when the Earth is full accreted (see Table 2) - based on uncertainties in the element partitioning models as listed in Table S4 of the supplementary information.

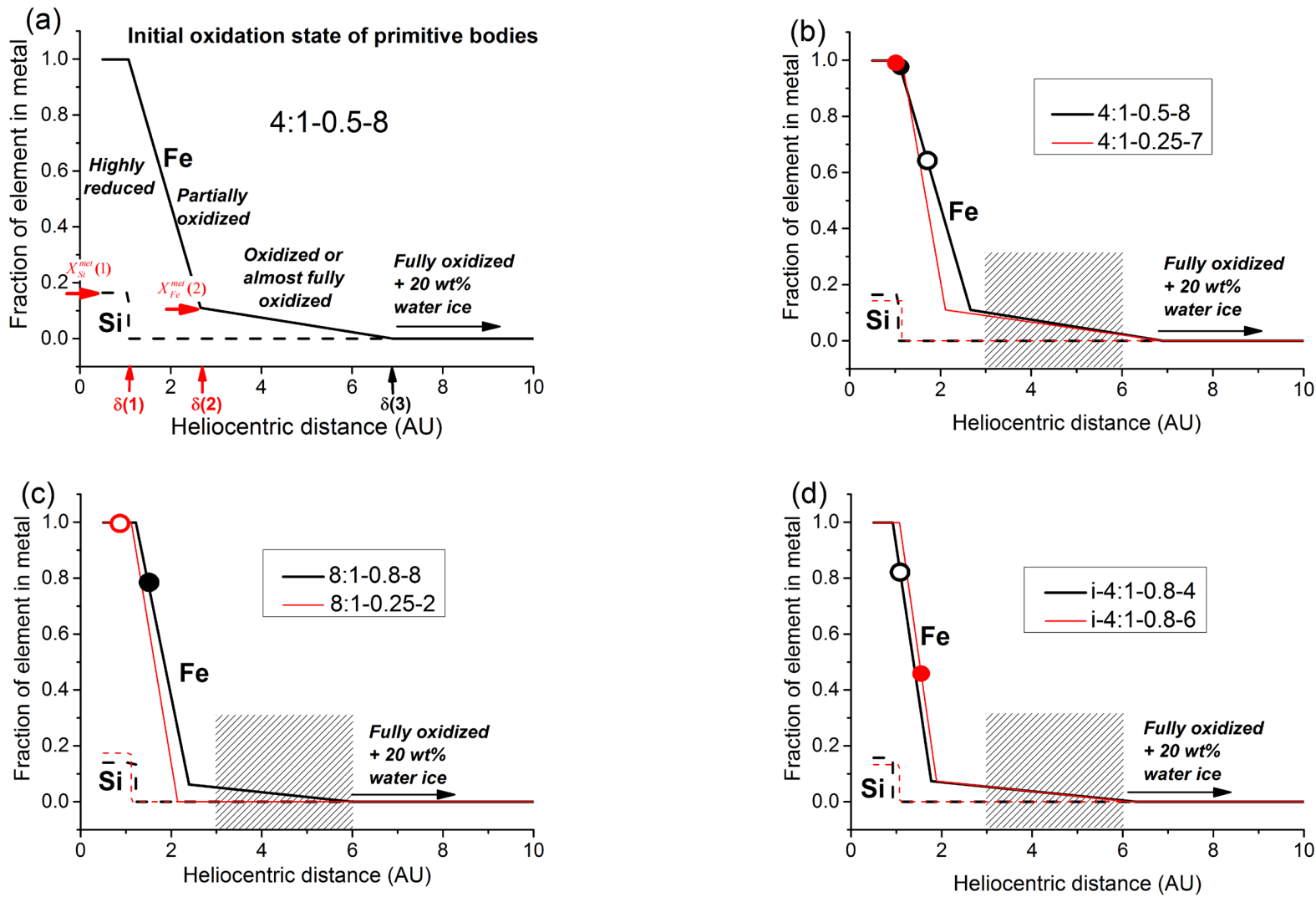

**Figure 6. (a)** Model of the compositions of primitive bodies (oxidation state and $H_2O$ content) as a function of their heliocentric distances of origin for simulation 4:1-0.5-8. The four parameters that have been refined in least squares regressions are indicated in red. The distance parameter $\delta(3)$ is not a fitting parameter but is adjusted to produce an Earth's mantle with a $H_2O$ concentration of ~1000 ppm (see text). **(b-d)** Compositional profiles derived for the six Grand Tack simulations (see Table 4 for values of fit parameters). The circular symbols show the original locations and oxidation states of the embryos that grow to form Mars-like planets (Fig. 1). The filled circles are for Mars-like planets that were almost fully accrete within 10 My, whereas the open symbols are for Mars-like planets that have much longer accretion times (see Fig. 2). Note that the original locations of the embryos that grow to form three Mars-like planets in simulation 4:1-0.25-7 are almost identical, as shown by the single red symbol in (b). The shaded rectangles in (b) to (d) show the region from 3-6 AU that is initially devoid of primitive bodies due to the accretion of Jupiter and Saturn.

**Fig. 7**

(a)

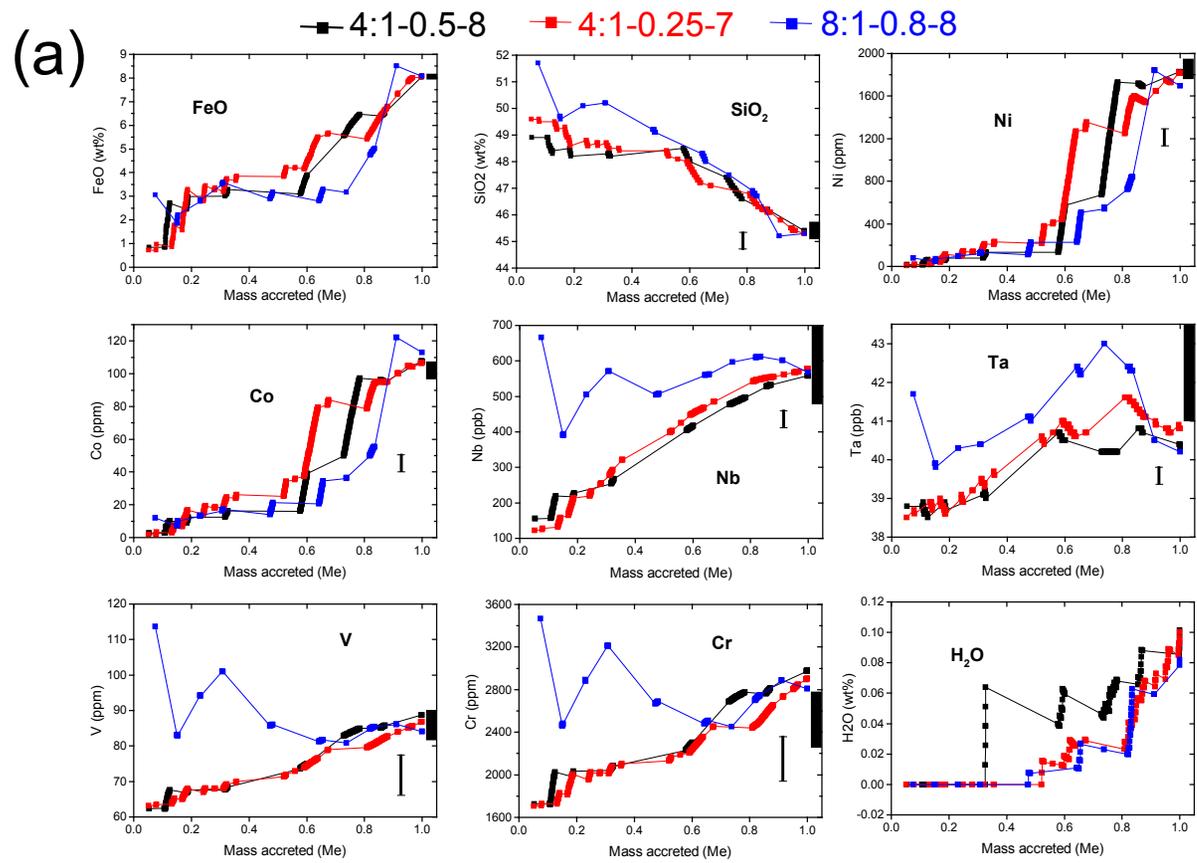

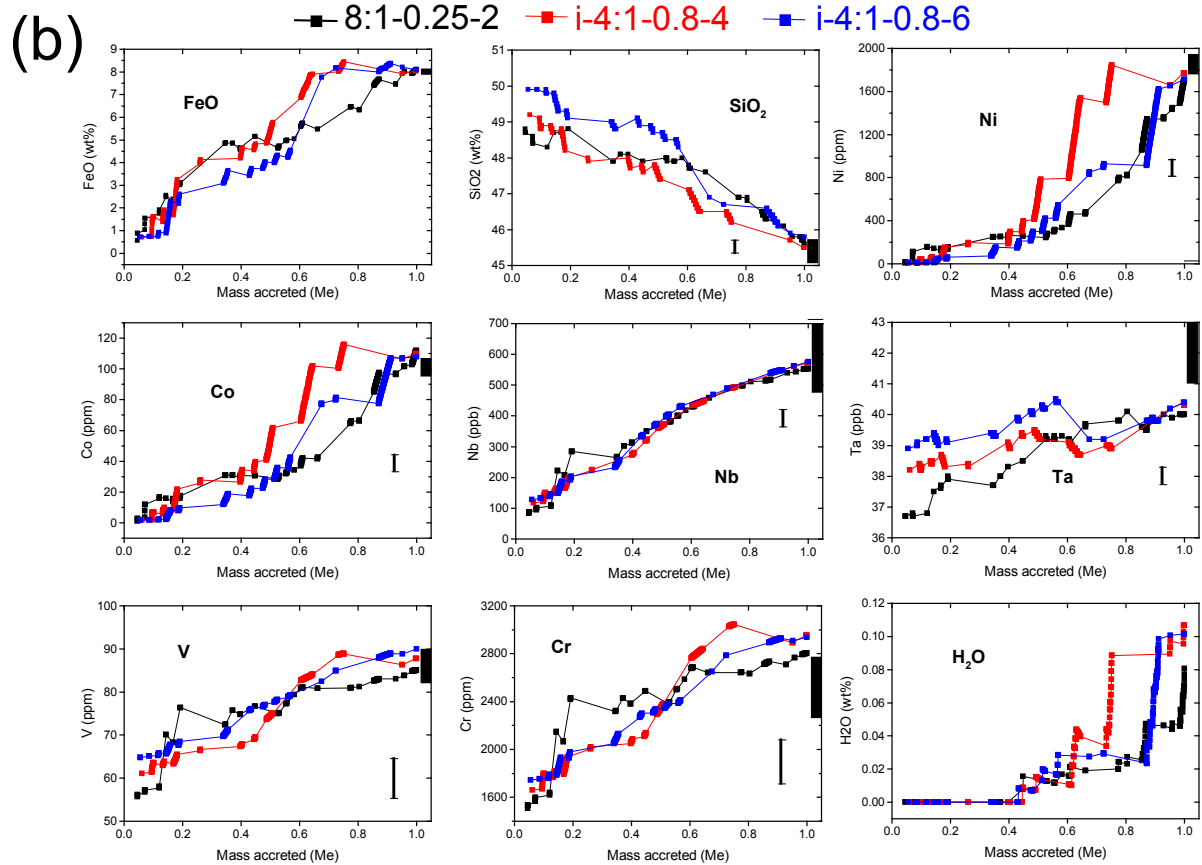

**Figure 7**. Evolution of element concentrations in the Earth's mantle as a function of mass accreted for the heterogeneous accretion/core formation model based on the composition-distance models of Fig. 6 (Tables 4 and 5). Results are shown for Grand Tack simulations 4:1-0.5-8, 4:1-0.25-7 and 8:1-0.8-8 **(a)**, and 8:1-0.25-2, i-4:1-0.8-4 and i-4:1-0.8-6 **(b)**. Estimated primitive mantle concentrations (Palme and O'Neill, 2013) are shown as a black bar on the right-hand axis of each graph. The error bars indicate the magnitude of the propagated uncertainties on the final concentrations when the Earth is full accreted based on uncertainties in the element partitioning models as listed in Table S4 of the Supplementary Data.

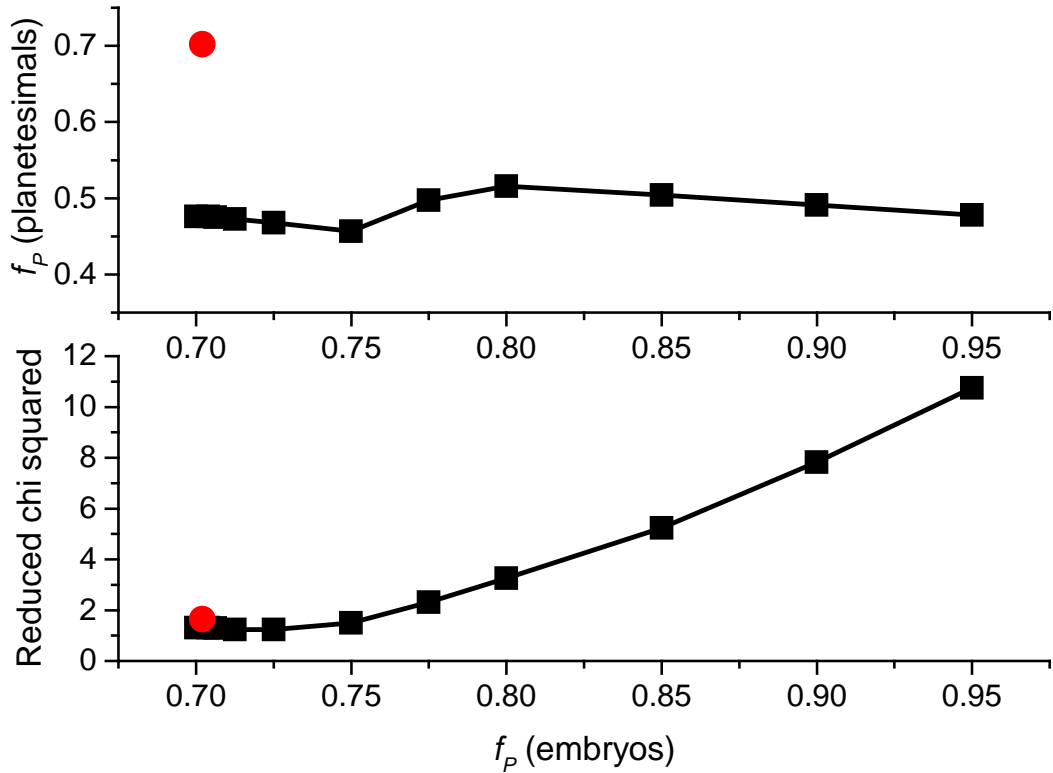

Figure 8. Results of least squares fitting when metal-silicate equilibration pressures (as given by $f_P$, Eq. 7) are assumed to be different for embryo and planetesimals impacts in simulation 4:1-0.25-7. Top: Refined values of $f_P$ for planetesimal impacts lie in the range 0.45-0.5 when $f_P$ for embryo impacts is fixed at values that range from 0.7 to 0.95. Bottom: Corresponding values of $\chi^2_{red}$: good fits are obtained when $f_P$ values for embryos lie in the range 0.7-0.75. The red symbols show the respective values when $f_P$ is identical for embryos and planetesimals.

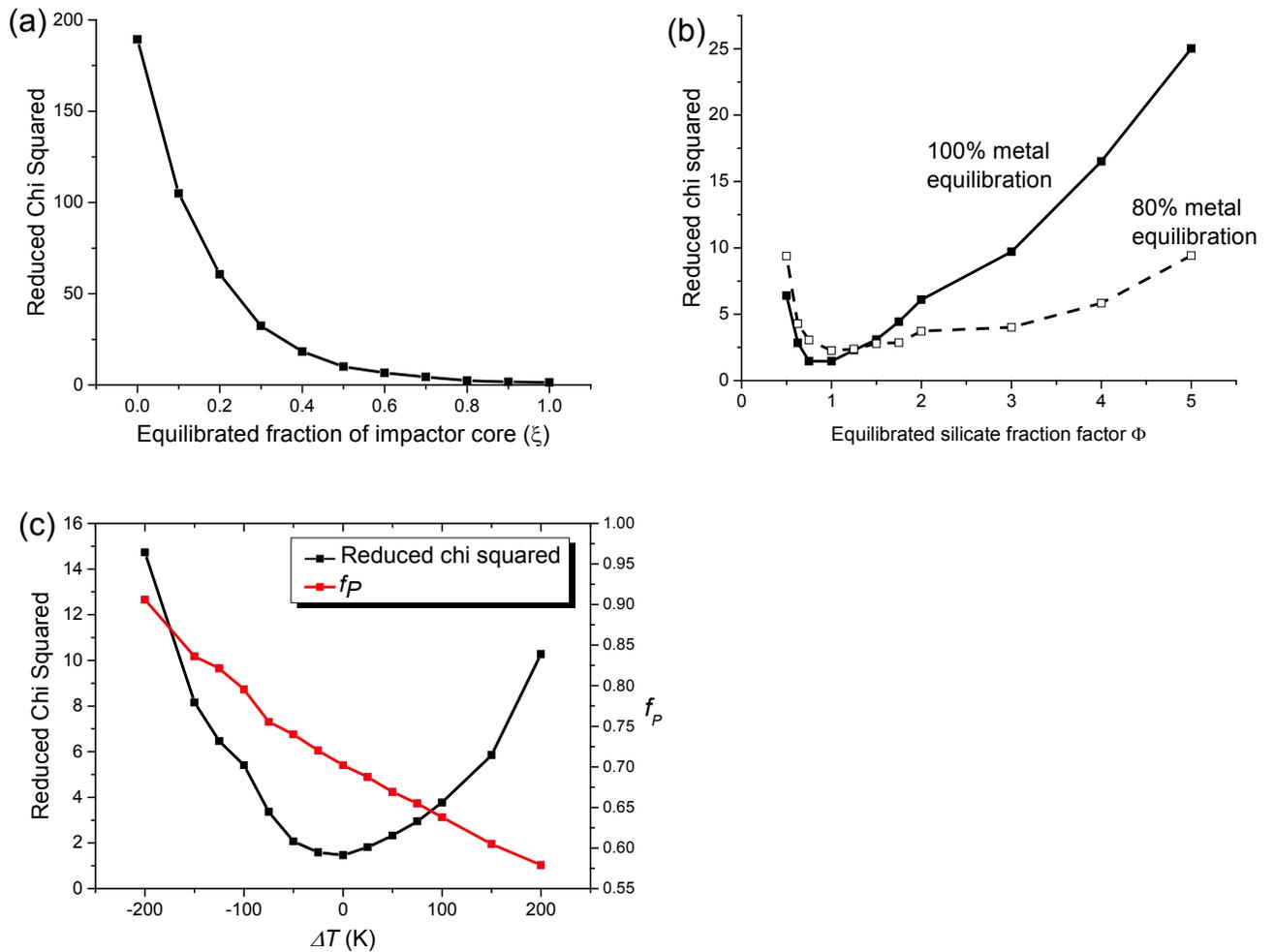

Figure 9. Effects of varying additional parameters on the quality of the least squares fit of the 4:1-0.25-7 heterogeneous accretion/core formation model (Fig. 7a and Table 5). (a) Incomplete equilibration of impactor cores results in poorer fits (increased chi squared), particularly when the equilibrated fraction of impactor cores is <0.6. (b) Effect of increasing or decreasing the fraction of silicate mantle that equilibrates with impactor cores. The fraction determined by Eq. 9 is changed by multiplying it by the factor $\Phi$. The solid line shows results with the metal of impactor cores equilibrating completely. The results with 80% metal equilibration show that $\Phi$ and the extent of metal equilibration are correlated. (c) The effect of adjusting the equilibration temperature model (Eq. 8) by $\Delta T$. The red line shows the effect on the refined value of $f_P$ (Eq. 7).

Table 1. Details of the six Grand Tack N-body accretion simulations.

| Simulation | Embryo source region (AU) | Planetesimal source* regions (AU) | Mass of embryos ($M_e$) | Mass of planetesimals* ($M_e$) | Number of embryos | Number of Planetesimals* | Mass adjustment** |
|---|---|---|---|---|---|---|---|
| 4:1-0.5-8 | 0.7 - 3.0 | 0.7 - 3.0<br>6.0  - 9.5 | 0.05 | $3.9 \times 10^{-4}$<br>$4.8 \times 10^{-5}$ | 87 | 2836<br>1563 | 1.06 |
| 4:1-0.25-7 | 0.7 - 3.0 | 0.7 - 3.0<br>6.0 - 9.5 | 0.025 | $3.9 \times 10^{-4}$<br>$4.8 \times 10^{-5}$ | 170 | 2783<br>1563 | 1.04 |
| 8:1-0.8-8 | 0.7 - 3.0 | 0.7 - 3.0<br>6.0 - 9.4 | 0.08 | $3.9 \times 10^{-4}$<br>$4.8 \times 10^{-5}$ | 68 | 1750<br>500 | 0.94 |
| 8:1-0.25-2 | 0.7 - 3.0 | 0.7 - 3.0<br>6.0 - 9.4 | 0.025 | $3.9 \times 10^{-4}$<br>$1.2 \times 10^{-4}$ | 213 | 1750<br>500 | 0.91 |
| i-4:1-0.8-4 | 0.5 - 3.0 | 0.5 - 3.0<br>6.0 - 9.4 | 0.02 – 0.08 | $3.9 \times 10^{-4}$<br>$1.2 \times 10^{-4}$ | 82 | 2260<br>500 | 1.01 |
| i-4:1-0.8-6 | 0.5 - 3.0 | 0.5 - 3.0<br>6.0 - 9.4 | 0.02 – 0.08 | $3.9 \times 10^{-4}$<br>$1.2 \times 10^{-4}$ | 82 | 2260<br>500 | 0.92 |

* Upper values are for the inner planetesimal disk and lower values are for the outer disk.

The region from 3 to 6 AU is cleared of planetesimals by the accretion of Jupiter and Saturn.

** The masses of all bodies were adjusted by this factor in order that each "Earth-like" planet had a final mass of exactly 1 $M_e$.

**Table 2**. Calculated mantle compositions and fit parameters for the Earth-like planets in the six Grand Tack simulations based on two homogeneous accretion models (Fig. 4a).

| Simulation: | | 4:1-0.5-8 | 4:1-0.25-7 | 8:1-0.8-8 | 8:1-0.25-2 | i-4:1-0.8-4 | i-4:1-0.8-6 |
|---|---|---|---|---|---|---|---|
| Heliocentric distance | | 0.969 AU | 0.907 AU | 1.18 AU | 0.936 AU | 0.79 AU | 0.88 AU |
| **(a) Homogeneous accretion – Best fit model** | | | | | | | |
| | **Earth PM** | | | | | | |
| SiO$_2$ | **45.40 (0.30)** | 47.4 (0.1) | 47.4 (0.1) | 47.3 (0.1) | 47.2 (0.1) | 47.2 (0.1) | 46.9 (0.1) |
| FeO | **8.10 (0.05)** | 8.11 (0.01) | 8.10 (0.01) | 8.12 (0.01) | 8.11 (0.01) | 8.11 (0.01) | 8.10 (0.01) |
| Ni ppm | **1860 (93)** | 1990 (373) | 2010 (491) | 1590 (360) | 1520 (452) | 1950 (404) | 1850 (555) |
| Co ppm | **102 (5)** | 110 (23) | 110 (28) | 94.4 (23) | 85.5 (26) | 110 (25) | 100 (30) |
| Nb ppb | **595 (119)** | 769 (17) | 770 (39) | 770 (19) | 774 (36) | 772 (18) | 778 (44) |
| Ta ppb | **43 (2)** | 41 (2) | 41 (1) | 41 (1) | 41 (1) | 41 (1) | 41 (1) |
| Nb/Ta | **14.0 (0.3)*** | 18.8 | 18.8 | 18.8 | 18.8 | 18.8 | 18.8 |
| V ppm | **86 (5)** | 121 (11) | 122 (10) | 121 (11) | 122 (10) | 121 (11) | 121 (11) |
| Cr ppm | **2520 (252)** | 4370 (383) | 4460 (320) | 4350 (384) | 4470 (317) | 4360 (378) | 4400 (318) |
| **Core mass fraction** | **0.32** | 0.29 | 0.28 | 0.30 | 0.29 | 0.30 | 0.31 |
| $f_p$ | | 0.74 | 0.72 | 0.72 | 0.73 | 0.75 | 0.80 |
| $X_{Fe}^{met}$ | | 0.85 | 0.85 | 0.84 | 0.84 | 0.85 | 0.84 |
| $\chi_{red}^2$ | | 70.3 | 71.5 | 71.0 | 73.5 | 69.0 | 67.5 |
| **(b) Homogeneous accretion – Core formation under oxidizing conditions** | | | | | | | |
| | **Earth PM** | | | | | | |
| SiO$_2$ | **45.40 (0.30)** | 42.9 (0.1) | 43.2 (0.1) | 42.4 (0.1) | 42.9 (0.1) | 42.4 (0.1) | 42.8 (0.1) |
| FeO | **8.10 (0.05)** | 17.7 (0.03) | 17.2 (0.01) | 18.0 (0.01) | 18.1 (0.01) | 17.3 (0.01) | 17.7 (0.01) |
| Ni ppm | **1860 (93)** | 3020 (469) | 3460 (509) | 2740 (550) | 2250 (400) | 3830 (735) | 2750 (474) |
| Co ppm | **102 (5)** | 195 (32) | 207 (34) | 189 (36) | 154 (27) | 222 (43) | 179 (32) |
| Nb ppb | **595 (119)** | 680 (54) | 681 (53) | 684 (53) | 675 (51) | 695 (65) | 682 (52) |
| Ta ppb | **43 (2)** | 36 (3) | 36 (1) | 36 (1) | 36 (1) | 37 (2) | 36 (1) |
| Nb/Ta | **14.0 (0.3)*** | 18.9 | 18.9 | 18.9 | 18.9 | 18.9 | 18.9 |
| V ppm | **86 (5)** | 113 (5) | 113 (5) | 112 (5) | 112 (5) | 112 (5) | 113 (5) |
| Cr ppm | **2520 (252)** | 4480 (163) | 4460 (177) | 4450 (157) | 4470 (151) | 4440 (170) | 4460 (163) |
| **Core mass fraction** | **0.32** | 0.21 | 0.22 | 0.22 | 0.21 | 0.23 | 0.22 |
| $f_p$ | | 0.74 | 0.66 | 0.78 | 0.64 | 1.0 | 0.71 |
| $X_{Fe}^{met}$ | | 0.55 | 0.55 | 0.55 | 0.55 | 0.55 | 0.55 |
| $\chi_{red}^2$ | | 5400 | 4937 | 5757 | 5801 | 5059 | 5346 |

**Earth PM**: Primitive Earth mantle composition from Palme and O'Neill (2013) listing elements on which least squares regressions are based.

Element concentrations are listed as wt% unless otherwise specified.

* Nb/Ta ratio from Münker et al. (2003).

Compositional uncertainties for the calculated Earth mantle compositions are based on log $K_D$ error propagations.

**Table 3**. Fitted parameters and results for heterogeneous accretion of the Earth- and Mars-like planets in the six Grand Tack simulations based on the stepped composition-distance model for primitive bodies of Fig. 4c.

| Simulation: | | 4:1-0.5-8 | 4:1-0.25-7 | 8:1-0.8-8 | 8:1-0.25-2 | i-4:1-0.8-4 | i-4:1-0.8-6 |
|---|---|---|---|---|---|---|---|
| $f_p$ | | 0.70 | 0.69 | 0.69 | 0.68 | 0.71 | 0.65 |
| $X_{Fe}^{met}(1)$ | | 0.999 | 0.999 | 0.999 | 0.999 | 0.999 | 0.999 |
| $X_{Si}^{met}(1)$ | | 0.106 | 0.125 | 0.113 | 0.118 | 0.126 | 0.137 |
| $X_{Fe}^{met}(2)$ | | 0.256 | 0.277 | 0.257 | 0.220 | 0.247 | 0.236 |
| $\delta(1)$ | | 1.72 AU | 1.53 AU | 1.72 AU | 1.48 AU | 1.10 AU | 1.37 AU |
| $\chi_{red}^2$ | | 2.0 | 0.82 | 2.2 | 3.0 | 1.5 | 2.2 |
| **Earth mantle compositions** | | | | | | | |
| | **Earth PM** | | | | | | |
| SiO₂ | **45.40 (0.30)** | 45.7 (0.1) | 45.3 (0.1) | 45.5 (0.1) | 45.3 (0.1) | 45.3 (0.1) | 45.2 (0.1) |
| FeO | **8.10 (0.05)** | 8.09 (0.01) | 8.09 (0.01) | 8.10 (0.01) | 8.11 (0.01) | 8.09 (0.01) | 8.08 (0.01) |
| Ni ppm | **1860 (93)** | 1790 (300) | 1810 (401) | 1650 (320) | 1570 (326) | 1740 (287) | 1690 (308) |
| Co ppm | **102 (5)** | 113 (19) | 107 (22) | 109 (19) | 103 (19) | 108 (18) | 110 (19) |
| Nb ppb | **595 (119)** | 567 (44) | 584 (40) | 575 (47) | 590 (31) | 590 (35) | 567 (32) |
| Ta ppb | **43 (2)** | 41 (1) | 41 (1) | 41 (1) | 41 (1) | 41 (1) | 41 (1) |
| Nb/Ta | **14.0 (0.3)\*** | 13.8 | 14.2 | 14.0 | 14.3 | 14.3 | 13.8 |
| V ppm | **86 (5)** | 86 (31) | 85 (31) | 83 (32) | 85 (33) | 83 (33) | 82 (33) |
| Cr ppm | **2520 (252)** | 2770 (469) | 2770 (444) | 2660 (470) | 2680 (477) | 2610 (473) | 2600 (467) |
| **Earth core compositions and mass fractions** | | | | | | | |
| Fe | | 82.2 | 81.3 | 81.8 | 80.9 | 81.1 | 81.0 |
| Ni | | 5.20 | 5.01 | 5.23 | 5.15 | 5.15 | 5.16 |
| Co | | 0.24 | 0.23 | 0.24 | 0.24 | 0.24 | 0.24 |
| O | | 3.47 | 3.62 | 3.39 | 4.14 | 3.96 | 3.76 |
| Si | | 8.16 | 9.06 | 8.52 | 8.80 | 8.78 | 9.02 |
| Nb ppb | | 539 | 526 | 518 | 489 | 486 | 540 |
| Ta ppb | | 4.1 | 4.6 | 4.3 | 4.1 | 4.4 | 5.1 |
| V ppm | | 124 | 131 | 129 | 125 | 130 | 131 |
| Cr wt% | | 0.76 | 0.78 | 0.77 | 0.76 | 0.77 | 0.78 |
| Core mass fraction | **0.32** | 0.309 | 0.306 | 0.314 | 0.316 | 0.317 | 0.318 |
| **Mars** | | | | | | | |
| Mantle FeO | **18.1 (1.0) \*** | a) 17.7 b) 3.05 | a) 5.84 b) 3.68 c) 5.53 | 2.18 | 11.0 | 0.23 | 28.7 |
| Core mass fraction | **0.15-0.23\*\*** | a) 0.21 b) 0.33 | a) 0.31 b) 0.33 c) 0.32 | 0.34 | 0.27 | 0.17 | 0.01 |

**Earth PM**: Primitive Earth mantle composition from Palme and O'Neill (2013) listing elements on which least squares regressions are based. Element concentrations are listed as wt% unless otherwise specified. Earth mantle Nb/Ta is ratio from Münker et al. (2003).

Compositional uncertainties for the calculated Earth mantle compositions are based on log $K_D$ error propagations.

\* Taylor (2013); \*\* Sohl and Spohn (1997); McSween (2003)

**Table 4**

Fitted parameter values and corresponding $\chi^2_{red}$ values based on the composition-distance model for primitive bodies of Fig. 6. Results are presented in Tables 5 to 7 and Fig. 7.

|  | 4:1-0.5-8 | 4:1-0.25-7 | 8:1-0.8-8 | 8:1-0.25-2 | i-4:1-0.8-4 | i-4:1-0.8-6 |
|---|---|---|---|---|---|---|
| $f_P$ | 0.72 | 0.70 | 0.66 | 0.58 | 0.70 | 0.65 |
| $X^{met}_{Fe}(1)$ * | 0.999 | 0.999 | 0.999 | 0.999 | 0.999 | 0.999 |
| $X^{met}_{Si}(1)$ | 0.164 | 0.143 | 0.140 | 0.174 | 0.158 | 0.133 |
| $X^{met}_{Fe}(2)$ | 0.11 | 0.11 | 0.064 | $5 \times 10^{-5}$ | 0.074 | 0.22 |
| $\delta(1)$ | 1.07 AU | 1.15 AU | 1.22 AU | 1.12 AU | 0.924 AU | 1.07 AU |
| $\delta(2)$ | 2.66 AU | 2.11 AU | 2.39 AU | 2.14 AU | 1.77 AU | 1.89 AU |
| $\delta(3)$ ** | 6.9 AU | 6.9 AU | 6.0 AU | 6.0 AU | 6.3 AU | 6.15 AU |
| $\chi^2_{red}$ | 2.3 | 1.6 | 3.5 | 3.6 | 2.9 | 3.8 |

* Value of $X^{met}_{Fe}(1)$ was kept constant in all least squares refinements.

** $\delta(3)$ was not a fitting parameter but was adjusted to give ~1000 ppm $H_2O$ in the Earth's mantle.

**Table 5**. Calculated final mantle and core compositions for the heterogeneous accretion of the Earth-like planets in the six Grand Tack simulations based on the composition-distance model for primitive bodies of Fig. 6. Parameter values are listed in Table 4.

| Simulation: | | 4:1-0.5-8 | 4:1-0.25-7 | 8:1-0.8-8 | 8:1-0.25-2 | i-4:1-0.8-4 | i-4:1-0.8-6 |
|---|---|---|---|---|---|---|---|
| **Heliocentric distance** | | 0.969 AU | 0.907 AU | 1.18 AU | 0.936 AU | 0.79 AU | 0.88 AU |
| **Mantle compositions** | | | | | | | |
| | **Earth PM** | | | | | | |
| SiO$_2$ | **45.40 (0.30)** | 45.4 (0.1) | 45.3 (0.1) | 45.5 (0.1) | 45.6 (0.1) | 45.5 (0.1) | 45.8 (0.1) |
| FeO | **8.10 (0.05)** | 8.09 (0.01) | 8.09 (0.01) | 8.10 (0.01) | 8.12 (0.01) | 8.08 (0.01) | 8.10 (0.01) |
| Ni ppm | **1860 (93)** | 1820 (321) | 1810 (404) | 1750 (265) | 1705 (222) | 1770 (295) | 1714 (294) |
| Co ppm | **102 (5)** | 108 (19) | 106 (23) | 115 (15) | 112 (14) | 110 (18) | 108 (19) |
| Nb ppb | **595 (119)** | 557 (35) | 578 (38) | 567 (42) | 553 (31) | 572 (28) | 576 (25) |
| Ta ppb | **43 (2)** | 40 (1) | 41 (1) | 41 (1) | 40 (1) | 40 (1) | 40 (1) |
| Nb/Ta | **14.0 (0.3)*** | 13.8 | 14.2 | 13.95 | 13.85 | 14.19 | 14.25 |
| V ppm | **86 (5)** | 89 (27) | 87 (29) | 84 (30) | 85 (28) | 88 (26) | 90 (28) |
| Cr ppm | **2520 (252)** | 2980 (426) | 2900 (415) | 2730 (452) | 2805 (433) | 2958 (424) | 2940 (449) |
| H$_2$O ppm | **1160 (232)** | 1010 | 1000 | 793 | 806 | 1066 | 1013 |
| Mass % accreted from H$_2$O-bearing bodies | | 0.38 % | 0.34 % | 0.26 % | 0.27 % | 0.37 % | 0.36 % |
| **Core compositions and mass fractions** | | | | | | | |
| Fe | | 81.7 | 81.3 | 82.4 | 82.9 | 81.6 | 82.3 |
| Ni | | 5.14 | 4.99 | 5.23 | 5.28 | 5.16 | 5.23 |
| Co | | 0.24 | 0.23 | 0.24 | 0.24 | 0.24 | 0.24 |
| O | | 3.59 | 3.71 | 2.89 | 2.58 | 3.85 | 3.81 |
| Si | | 8.65 | 9.03 | 8.51 | 8.23 | 8.40 | 7.73 |
| Nb ppb | | 557 | 532 | 533 | 562 | 514 | 500 |
| Ta ppb | | 5.7 | 5.1 | 5.0 | 6.2 | 5.5 | 4.6 |
| V ppm | | 118 | 127 | 128 | 125 | 118 | 113 |
| Cr wt% | | 0.70 | 0.74 | 0.75 | 0.74 | 0.69 | 0.70 |
| H ppm | | 58 | 28 | 14 | 22 | 34 | 40 |
| Core mass fraction | **0.32** | 0.312 | 0.306 | 0.313 | 0.311 | 0.314 | 0.311 |

**Earth PM**: Primitive Earth mantle composition from Palme and O'Neill (2013) listing elements on which least squares regressions are based.

Element concentrations are listed as wt% unless otherwise specified.

* Nb/Ta ratio from Münker et al. (2003).

Compositional uncertainties for the calculated Earth mantle compositions are based on log $K_D$ error propagations.

**Table 6**. Calculated final mantle and core compositions for the Venus-like planets in the six Grand Tack simulations based on the composition-distance model for primitive bodies of Fig. 6. Parameter values are listed in Table 4.

| Simulation: | 4:1-0.5-8 | 4:1-0.25-7 | 8:1-0.8-8 | 8:1-0.25-2 | i-4:1-0.8-4 | i-4:1-0.8-6 |
|---|---|---|---|---|---|---|
| **Heliocentric distance** | 0.623 AU | 0.594 AU | 0.650 AU | 0.608 AU | 0.530 AU | 0.572 AU |
| **Mass** | 0.979 Me | 0.745 Me | 0.987 Me | 0.766 Me | 0.701 Me | 0.808 Me |
| **Mantle compositions** | | | | | | |
| SiO$_2$ | 45.7 | 46.0 | 47.2 | 46.5 | 46.1 | 47.1 |
| FeO | 7.38 | 8.25 | 4.52 | 7.13 | 7.92 | 5.92 |
| Ni ppm | 1946 | 2482 | 761 | 1072 | 1863 | 1796 |
| Co ppm | 108 | 145 | 52 | 78 | 116 | 98 |
| Nb ppb | 616 | 530 | 569 | 413 | 494 | 484 |
| Ta ppb | 41 | 40 | 42 | 38 | 39 | 41 |
| Nb/Ta | 15.0 | 13.39 | 13.51 | 10.77 | 12.54 | 11.93 |
| V ppm | 92 | 88 | 84 | 78 | 84 | 79 |
| Cr ppm | 3134 | 2880 | 2626 | 2571 | 2741 | 2468 |
| H$_2$O ppm | 824 | 1190 | 884 | 1515 | 580 | 773 |
| Mass % accreted from H$_2$O-bearing bodies | 0.27 % | 0.31 % | 0.29 % | 0.50 % | 0.20 % | 0.27 |
| **Core compositions and mass fractions** | | | | | | |
| Fe | 85.3 | 84.4 | 83.0 | 85.2 | 84.6 | 85.1 |
| Ni | 4.88 | 5.21 | 5.13 | 5.44 | 5.33 | 5.15 |
| Co | 0.23 | 0.24 | 0.24 | 0.25 | 0.24 | 0.24 |
| O | 3.14 | 2.08 | 2.66 | 0.73 | 1.47 | 1.27 |
| Si | 9.11 | 7.29 | 8.26 | 7.57 | 7.61 | 7.45 |
| Nb ppb | 478 | 607 | 550 | 879 | 700 | 720 |
| Ta ppb | 5.2 | 5.7 | 5.4 | 10 | 7.2 | 6.3 |
| V ppm | 169 | 118 | 129 | 142 | 130 | 139 |
| Cr wt% | 0.73 | 0.73 | 0.76 | 0.79 | 0.76 | 0.80 |
| H ppm | 13 | 91 | 18 | 36 | 17 | 33 |
| Core mass fraction | 0.30 | 0.30 | 0.33 | 0.31 | 0.31 | 0.32 |

Element concentrations are listed as wt% unless otherwise specified.

**Table 7**. Calculated mantle and core compositions for Mars-like planets (with accretion times of <10 My) based on the composition-distance model for primitive bodies of Fig. 6. Parameter values are listed in Table 4.

| Simulation: | | 4:1-0.5-8 "Mars 2" | 4:1-0.25-7 "Mars-1" | 4:1-0.25-7 "Mars-3" | 8:1-0.8-8 | i-4:1-0.8-6 |
|---|---|---|---|---|---|---|
| **HD of origin:**<br>**Final HD:**<br>**Mass:** | **1.52 AU**<br>**0.107 M$_e$** | **1.14 AU**<br>**1.79 AU**<br>**0.064 M$_e$** | **1.13 AU**<br>**1.31 AU**<br>**0.035 M$_e$** | **1.16 AU**<br>**1.85 AU**<br>**0.032 M$_e$** | **1.53 AU**<br>**1.76 AU**<br>**0.081 M$_e$** | **1.58 AU**<br>**1.90 AU**<br>**0.064 M$_e$** |
| **Martian mantle compositions** | | | | | | |
| | **Taylor (2013)** | | | | | |
| SiO$_2$ | 43.7 (1.0) | 50.5 | 47.3 | 49.5 | 47.2 | 42.3 |
| FeO | 18.1 (1.0) | 4.94 | 6.71 | 6.80 | 11.8 | 20.2 |
| Ni ppm | 330 (109) | 102 | 170 | 168 | 331 | 568 |
| Co ppm | 71 (25) | 17.3 | 29.4 | 28.5 | 51.2 | 92 |
| Nb ppb | 501 (7) | 621 | 258 | 569 | 696 | 644 |
| Ta ppb | 27.2 (1) | 40 | 37 | 39 | 38 | 34 |
| Nb/Ta | 19.9 (0.06)* | 15.4 | 7.0 | 14.5 | 18.4 | 18.8 |
| V ppm | 60-105 (R&C) | 112 | 75 | 109 | 121 | 113 |
| Cr ppm | 4990 (420) | 3534 | 2437 | 3415 | 4651 | 4653 |
| H$_2$O ppm | | 1786 | 2225 | 4769 | 399 | 0 |
| Mass % accreted from H$_2$O-bearing bodies | | 0.57% | 0.68% | 1.51% | 0.13% | 0.0% |
| **Core compositions and mass fractions** | | | | | | |
| Fe | | 92.1 | 86.7 | 92.0 | 92.0 | 89.8 |
| Ni | | 5.9 | 5.7 | 6.1 | 7.0 | 9.3 |
| Co | | 0.27 | 0.26 | 0.28 | 0.32 | 0.41 |
| O | | 0.07 | 0.02 | 0.03 | 0.29 | 0.38 |
| Si | | 1.2 | 6.48 | 0.98 | 0.15 | 0.02 |
| Nb ppb | | 371 | 123 | 467 | 54.4 | 5.5 |
| Ta ppb | | 1.5 | 12.8 | 1.3 | 0.14 | 0.01 |
| V ppm | | 59 | 149 | 60 | 13 | 3.4 |
| Cr wt% | | 0.57 | 0.82 | 0.59 | 0.24 | 0.11 |
| H ppm | | 13 | 5.9 | 2.4 | 0 | 0 |
| Core mass fraction | 0.15-0.23** | 0.29 | 0.31 | 0.28 | 0.24 | 0.18 |

HD: Heliocentric distance
Element concentrations are listed as wt% unless otherwise specified.
Estimated Martian bulk silicate composition from Taylor (2013); R&C: Righter & Chabot (2011)
* Nb/Ta  ratio from Münker et al. (2003)
** Martian core mass fraction from Sohl and Spohn (1997) and McSween (2003)